\documentclass[10pt,journal,compsoc]{IEEEtran}
\IEEEoverridecommandlockouts
\usepackage{cite}
\usepackage{amsmath,amssymb,amsfonts}
\usepackage{algorithm}
\usepackage{algorithmic}
\usepackage{graphicx}
\usepackage{textcomp}
\usepackage{xcolor}
\usepackage{optidef} 
\usepackage{verbatim}
\usepackage{todonotes}
\usepackage{caption}
\expandafter\def\expandafter\normalsize\expandafter{%
    \normalsize%
    \setlength\abovedisplayskip{0pt}%
    \setlength\belowdisplayskip{8pt}%
    \setlength\abovedisplayshortskip{-1pt}%
    \setlength\belowdisplayshortskip{2pt}%
}
\usepackage{subcaption}
\usepackage{todonotes}
\usepackage{titlesec}         
\DeclareMathSizes{10}{9}{6}{5}
\DeclareMathSizes{10}{9}{6}{5}

\begin{document}
\title{\Large{Design Optimization of NOMA Aided Multi-STAR-RIS for Indoor Environments:\\A Convex Approximation Imitated Reinforcement Learning Approach}}
\author{Yu~Min~Park,~\IEEEmembership{Student~Member,~IEEE},~Sheikh~Salman~Hassan,~\IEEEmembership{Member,~IEEE},~Yan~Kyaw~Tun,~\IEEEmembership{Member,~IEEE},~Eui-Nam~Huh,~\IEEEmembership{Member,~IEEE},~Walid~Saad,~\IEEEmembership{Fellow,~IEEE},~and~Choong~Seon~Hong,~\IEEEmembership{Fellow,~IEEE}

\IEEEcompsocitemizethanks{ \IEEEcompsocthanksitem Yu Min Park, Sheikh Salman Hassan, Eui-Nam Huh, and Choong Seon Hong are with the Department of Computer Science and Engineering, Kyung Hee University,  Yongin-si, Gyeonggi-do 17104, Rep. of Korea, e-mails:{\{yumin0906, salman0335, johnhuh, cshong\}@khu.ac.kr}.}

\IEEEcompsocitemizethanks{ \IEEEcompsocthanksitem Yan Kyaw Tun is with the Department of Electronic Systems, Aalborg University, A. C. Meyers Vænge 15, 2450 København, e-mail: ykt@es.aau.dk.}

\IEEEcompsocitemizethanks{ \IEEEcompsocthanksitem Walid Saad is with the Bradley Department of Electrical and Computer Engineering, Virginia Tech, VA, 24061, USA. Email: walids@vt.edu.}}

\markboth{Journal of \LaTeX\ Class Files,~Vol.~14, No.~8, August~2015}
{Shell 
\MakeLowercase{\textit{et al.}}: Bare Demo of IEEEtran.cls for IEEE Journals}

\IEEEtitleabstractindextext{\begin{abstract}
Non-orthogonal multiple access (NOMA) enables multiple users to share the same frequency band, and simultaneously transmitting and reflecting reconfigurable intelligent surface (STAR-RIS) provides 360-degree full-space coverage, optimizing both transmission and reflection for improved network performance and dynamic control of the indoor environment. However, deploying STAR-RIS indoors presents challenges in interference mitigation, power consumption, and real-time configuration. In this work, a novel network architecture utilizing multiple access points (APs), STAR-RISs, and NOMA is proposed for indoor communication. To address these, we formulate an optimization problem involving user assignment, access point (AP) beamforming, and STAR-RIS phase control. A decomposition approach is used to solve the complex problem efficiently, employing a many-to-one matching algorithm for user-AP assignment and K-means clustering for resource management. Additionally, multi-agent deep reinforcement learning (MADRL) is leveraged to optimize the control of the STAR-RIS. Within the proposed MADRL framework, a novel approach is introduced in which each decision variable acts as an independent agent, enabling collaborative learning and decision making. The MADRL framework is enhanced by incorporating convex approximation (CA), which accelerates policy learning through suboptimal solutions from successive convex approximation (SCA), leading to faster adaptation and convergence. Simulations demonstrate significant improvements in network utility compared to baseline approaches.
\end{abstract}
}
\maketitle

\IEEEdisplaynontitleabstractindextext

\IEEEpeerreviewmaketitle

\ifCLASSOPTIONcompsoc
\IEEEraisesectionheading{\section{Introduction}\label{sec:introduction}}
\else

\section{Introduction}
\fi
\IEEEPARstart{R}{econfigurable} Intelligent Surfaces (RISs) emerged as a promising technology in wireless communication systems \cite{RIS_directions}. RISs, also known as intelligent reflecting surfaces (IRSs) or metasurfaces, consist of numerous sub-wavelength-sized elements that can dynamically manipulate the electromagnetic waves impinging upon them. These surfaces enable the control and optimization of wireless channels by adaptively modifying the phase, amplitude, and polarization of the incident waves \cite{RIS_lit_1}. In recent years, RISs have garnered significant attention from the research community due to their potential to enhance the performance of wireless networks in terms of coverage, capacity, and energy efficiency. By intelligently manipulating the propagation environment, RISs can mitigate path loss (both mmWave \cite{RIS_lit_2} and THz \cite{chaccour2024joint}), eliminate interference, and improve the signal-to-noise ratio. A RIS-aided hybrid wireless network comprising both active and passive components will be highly promising to achieve sustainable capacity growth cost-effectively in the future \cite{RIS_lit_4}. Different elements of a RIS can independently reflect the incident signal by controlling its amplitude and/or phase and thereby collaboratively achieve fine-grained 3D passive beamforming for directional signal enhancement or nulling \cite{RIS_lit_5}. A RIS-assisted MEC-enabled UAV system in the context of 6G THz communication networks is analyzed to optimize UAV computation power, RIS phase shifting, and THz sub-band allocation to minimize network latency \cite{RIS_lit_6}. RISs offer exciting possibilities for wireless networks but require significant research advancements. Key challenges include modeling complex channels, optimizing RIS settings, practical considerations, and security concerns. Addressing these challenges is necessary to unlock the full potential of RISs for enhancing future wireless communications.

\begin{figure}[t]
\centering
\includegraphics[width=\linewidth]{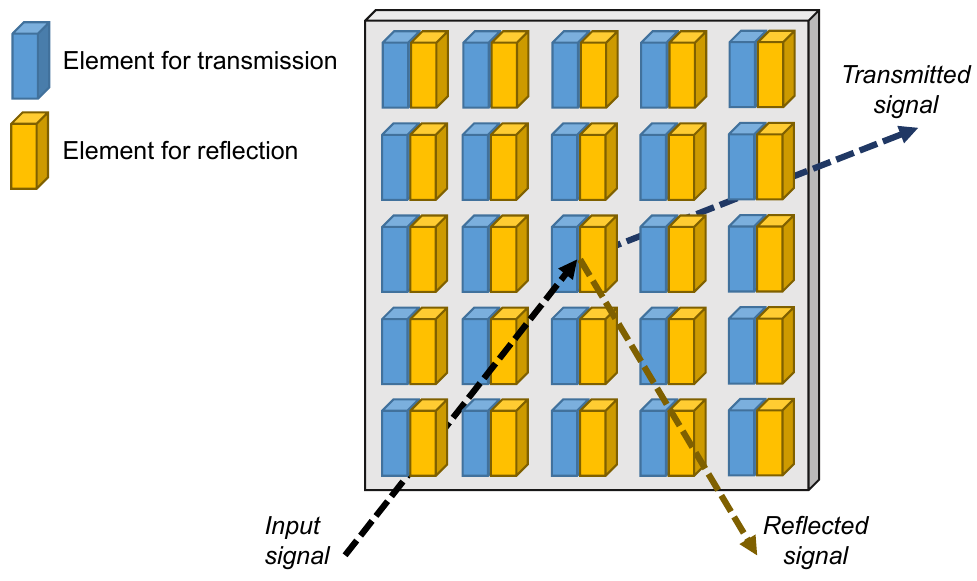}
\caption{STAR-RISs architecture.}
\label{star_ris}
\end{figure}

As shown in Fig. \ref{star_ris}, simultaneously transmitting and reflecting (STAR)-RISs is a novel concept that has been investigated in the literature \cite{STAR_RIS_lit_1}. The incident wireless signal is divided into transmitted and reflected signals passing into both sides of the space surrounding the surface, thus facilitating a full-space manipulation of signal propagation \cite{STAR_RIS_lit_2}. This dual functionality enables the surfaces to simultaneously enhance the received signal power and transmit additional information, thereby expanding the capabilities of traditional RISs \cite{STAR_RIS_lit_3}. \textcolor{black}{By intelligently manipulating the phase and amplitude of both the reflected and transmitted signals, STAR-RISs can achieve improved coverage, increased spectral efficiency, and enhanced communication reliability \cite{STAR_RIS_lit_4}.} The ergodic rates increase with the number of STAR-RIS elements \cite{STAR_RIS_lit_5}. This research investigates the application of multi-STAR-RIS technology in indoor environments with multiple users. We propose a novel and computationally efficient solution mechanism for network management in this complex scenario. This mechanism leverages a novel approach where imitated learned input decision variables are used for each agent. Multiple deep reinforcement learning (DRL) agents are then employed, each responsible for optimizing a specific decision variable based on its corresponding constraints. Our key contributions include:
\begin{itemize}
    \item We propose a novel network architecture for indoor environment wireless communication where multiple access points provide services to multiple users with the aid of multiple STAR-RIS.
    \item Drawing upon the proposed system architecture, we devise an optimization problem to address the user assignment, AP active beamforming, as well as amplitude and phase shift control of STAR-RISs.
    \item  To address the problem mentioned above, we treat each decision variable as an individual agent within the framework of DRL, as the constraints associated with each variable differ and do not yield optimal rewards for each respective agent.
    \item Additionally, we present a novel solution approach that incorporates learned (sub-optimal) values, which are imitated with the help of the convex approximation (CA) as referred to by the agent. This approach allows for swift learning of the environment's dynamics owing to the advantageous initialization of these variables. 
    \item Our simulation results were compared with conventional reinforcement learning methods, numerical optimization, and metaheuristic techniques. In the same learning conditions and test case, the network utility performs 15$\%$ higher than the baselines.
\end{itemize}

This paper is organized as follows. Section~\ref{literature} provides a comprehensive review of relevant prior research. Section~\ref{sys_model} introduces the system model and the formulated optimization problem.  Section~\ref{sol_app} details the proposed solution algorithm. The simulation setup and results are presented in Section~\ref{simul}. Finally, Section~\ref{conc} concludes the paper by summarizing our key findings and outlining potential future directions. 

\section{Related works}
\label{literature}
Few studies have addressed the joint management of phase shifts of reflection and transmission, resource allocation, and transmission techniques in indoor STAR-RIS network environments. The work \cite{STAR_RIS_lit_6} introduces an overarching optimization framework to address recent advancements in Star-RIS technology. This framework offers remarkable adaptability and demonstrably achieves optimal performance across various application contexts. Moreover, in \cite{STAR_RIS_lit_7}, Authors formulate a problem that maximizes the sum rate by jointly optimizing the trajectory, active beamforming of the UAV, and passive transmission/reflection beamforming of the STAR-RIS. The work in \cite{new_STAR-RIS_1} investigates the application of STAR-RIS in over-the-air computation (AirComp) systems. To overcome the limitations of traditional RISs and improve computation accuracy across a wider area, a joint beamforming design is proposed that optimizes transmit power, reflection/transmission at the STAR-RIS, and reception at the fusion center. Authors in \cite{new_STAR-RIS_2} studied NOMA communication systems aided by STAR-RIS. It proposes a joint optimization approach for maximizing user data rates, considering beamforming strategies at both the base station and the STAR-RIS. The work in \cite{new_STAR-RIS_3} addresses the limitations of prior RIS research by proposing a cooperative system combining double RIS and STAR-RIS technologies for enhanced coverage and performance in a massive MIMO setup under fading conditions. It introduces efficient channel estimation and low-complexity optimization for maximizing spectral efficiency. Moreover, the work in \cite{new_STAR-RIS_4} explores covert communication using a multi-antenna transmitter, a friendly jammer receiver, and a STAR-RIS. It proposes an optimization approach to maximize covert data rates while ensuring covertness and user quality of service. 

Non-orthogonal multiple access (NOMA) is a promising technique for next-generation networks, allowing multiple users to share the same band simultaneously through power level differentiation \cite{NOMA_lit_1}. Downlink NOMA with successive interference cancellation (SIC) offers improved capacity and cell-edge user performance in both frequency-selective and non-selective fading channels \cite{NOMA_lit_2}. The work \cite{NOMA_STAR_RIS_LIT_1} explores NOMA in conjunction with STAR-RISs. Authors investigate a coupled transmission and reflection phase-shift model for passive STAR-RISs, enabling power consumption minimization for both NOMA and Orthogonal Multiple Access where the joint optimization of amplitude and phase-shift coefficients for transmission and reflection is performed under user rate constraints. Research work in \cite{NOMA_STAR_RIS_LIT_2} explores suboptimal algorithms for maximizing the achievable sum rate in STAR-RIS-NOMA systems, including decoding order, power allocation, and beamforming optimization. Furthermore, studies demonstrate the performance benefits of STAR-RIS-NOMA compared to traditional systems, particularly in Rician fading channels \cite{NOMA_STAR_RIS_LIT_3}. Spectral efficiency improvements are explored through the application of Index Modulation (IM) in STAR-RIS-NOMA systems \cite{NOMA_STAR_RIS_LIT_4}. This approach allows for additional information transmission by assigning subsurfaces to users based on a predefined pattern. Finally, the analysis extends to short-packet communications (SPC) with NOMA and STAR-RISs, investigating the impact on common block length minimization under Rician fading and energy splitting protocols in scenarios with both continuous and discrete phase shifts \cite{NOMA_STAR_RIS_LIT_5}.

DRL is gaining traction for its ability to optimize network performance through learning from interactions within dynamic network environments \cite{DRL_LIT_2, DRL_LIT_1, TMC_own}. The work in \cite{DRL_STAR_RIS_LIT_1} explores DRL for optimizing communication systems aided by STAR-RISs, which formulates a power-minimizing problem for a multi-user MISO system with coupled phase-shift constraints, proposing the hybrid deep deterministic policy gradient (DDPG) algorithm and the joint DDPG \& deep-Q network (DDPG-DQN) algorithms to address it. Another study introduces MADRL-based JTORA for joint task offloading and resource allocation in NOMA-assisted MEC systems with STAR-RIS, aiming to improve communication quality under mode-switching protocols \cite{DRL_STAR_RIS_LIT_2}. Furthermore, DRL is being explored to mitigate challenges in robot communications. One approach utilizes Federated DRL (F-DRL) to address signal blockage and trajectory design for robots operating in dynamic environments, combining NOMA with RIS for improved connectivity \cite{DRL_NOMA_RIS_LIT_1}. Additionally, the research explores DRL with MAPPO for optimizing RIS deployment in satellite-based communication systems to overcome long-distance signal attenuation \cite{drl_ris_new}.

We highlight the challenges and unexplored research avenues about STAR-RISs. In particular, attention is drawn to critical areas requiring attention, including power allocation, interference management, and practical implementation considerations. Furthermore, the existing literature lacks a comprehensive exploration of the potential of Multi-STAR-RIS-assisted networking for indoor communication. Prior research studies have not collectively addressed the optimization of network utility in indoor environments encompassing multiple simultaneous users and multiple STAR-RISs. The deployment of Multi-STAR-RIS in indoor environments is justified by its ability to enhance signal coverage and quality, particularly in complex settings with obstacles like walls and furniture. It improves network capacity in high-density areas by reducing interference and directing signals more efficiently, while also being energy-efficient compared to traditional methods. Furthermore, Multi-STAR-RIS is a cost-effective solution for enhancing existing infrastructure and is critical for supporting future wireless technologies like 6G, which will rely on higher frequency bands that are more susceptible to signal degradation. In addition, previous studies have typically fed raw environmental data directly into the agent, which requires extensive computational time to achieve threshold convergence or the desired level of optimality. In contrast, we propose a novel approach where imitated learned input values are provided to the DRL agents, which enables swift convergence within the intricate network environment. 

\section{System Model}
\label{sys_model}
\begin{figure}[t]
\centering\includegraphics[width=\columnwidth]{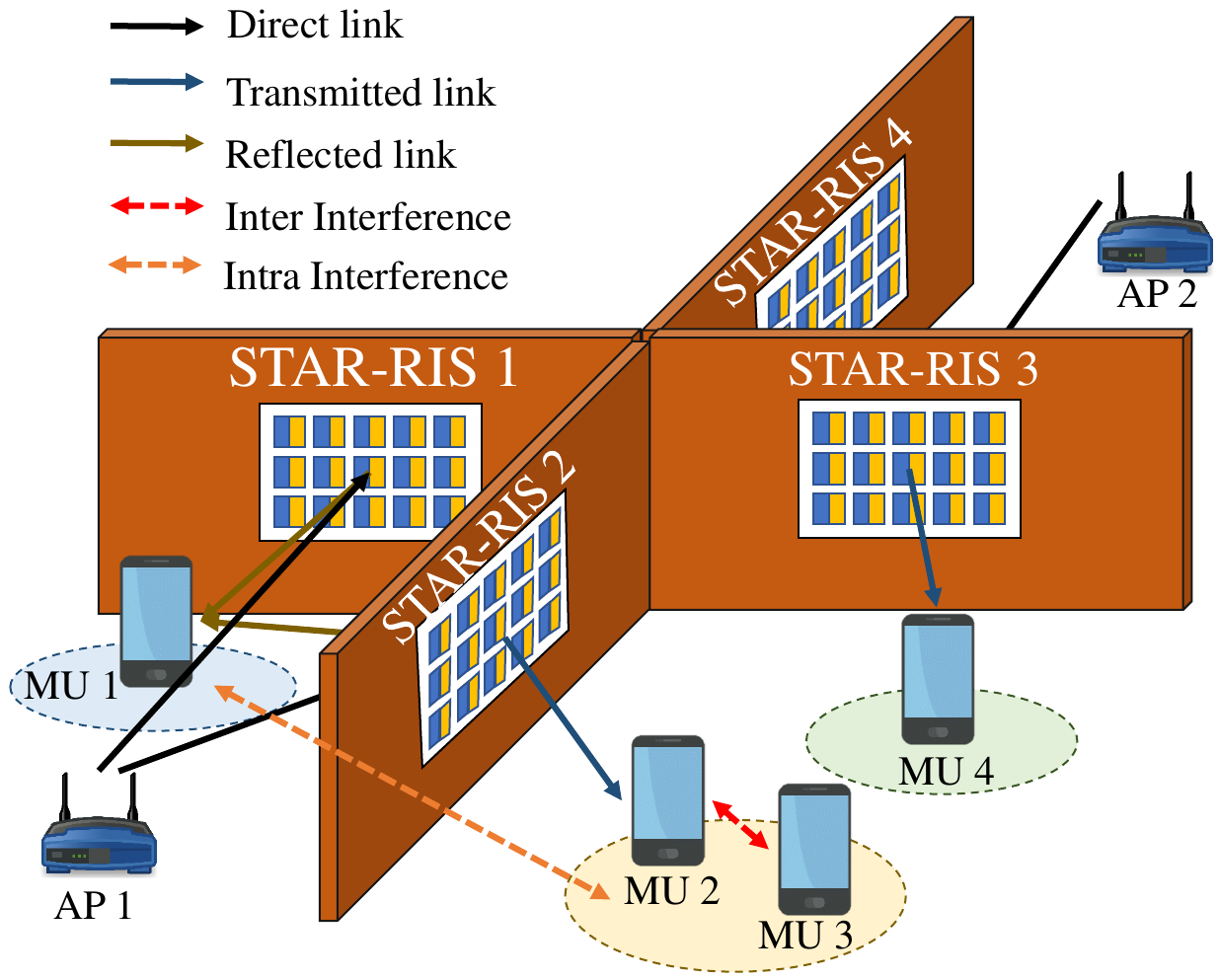}
\caption{NOMA aided Multi-STAR-RIS indoor networks.}
\label{system_model}
\end{figure}

\subsection{Network Model}
As illustrated in Fig.~\ref{system_model}, we consider a downlink communication with Multi-STAR-RIS aided NOMA in indoor environments, which consists of a set $\mathcal{B}$ of $B$ access points (APs) with $N_{b}$ antenna, a set $\mathcal{L}$ of $L$ STAR-RISs and a set $\mathcal{U}$ of $U$ user equipments (UEs) with a single-antenna. \textcolor{black}{In this work, we assume that APs operate on different frequency bands. As a result, there is no inter-cell interference between APs.} We also consider STAR-RISs with $M=M_{h}M_{v}$ elements, where $M_{h}$ and $M_{v}$ denote the number of elements along the vertical and the horizontal, respectively. The locations of AP $b \in \mathcal{B}$, center of STAR-RIS $l \in \mathcal{L}$ and UE $u \in \mathcal{U}$ are ${p}_{b}=\left[{x}_{b},{y}_{b},h_{b}\right]^{T}$, ${p}_{l}=\left[{x}_{l},{y}_{l},h_{l}\right]^{T}$ and ${p}_{u}=\left[{x}_{u},{y}_{u}\right]^{T}$. As shown in Fig.~\ref{system_model}, we assume that the environment under consideration consists of several rooms, and some of the walls of the rooms are composed of START-RIS. Downlink communication between APs and UEs in different rooms is possible through walls composed of STAR-RISs. In this paper, we assume that perfect channel state information (CSI) is available at the AP to investigate the performance gain of STAR Multi-RIS. The STAR-RIS adopts an energy splitting (ES) protocol, where each element can operate simultaneous transmission and reflection modes. For given transmission and reflection amplitude coefficients, the signals incident upon each element are split into transmitted and reflected signals having different energy. In a practical implementation, the amplitude and phase shift coefficients of each element for transmission and reflection will be jointly optimized for achieving diverse design objectives in wireless networks.

\subsection{Communication Model \& Link Analysis}
As shown in Fig.~\ref{system_model}, in our system model, the reflection and transmission surfaces of STAR-RIS are considered differently depending on the location of the APs. Thus, we assume that clockwise surfaces are forward side and otherwise backward side. Therefore, the forward and backward side passive beamforming vectors of STAR-RIS $l$ are given by: 
\begin{equation}
\boldsymbol{\Phi}^{F}_{l} = \textrm{diag}\left\{\sqrt{\beta^{F}_{l_1}} e^{j\theta^{F}_{l_1}}, \sqrt{\beta^{F}_{l_2}} e^{j\theta^{F}_{l_2}}, ..., \sqrt{\beta^{F}_{l_M}} e^{j\theta^{F}_{l_M}} \right\},\label{eq_Phi_t_l}
\end{equation}
\begin{equation}
\boldsymbol{\Phi}^{B}_{l} = \textrm{diag}\left\{\sqrt{\beta^{B}_{l_1}} e^{j\theta^{B}_{l_1}}, \sqrt{\beta^{B}_{l_2}} e^{j\theta^{B}_{l_2}}, ..., \sqrt{\beta^{B}_{l_M}} e^{j\theta^{B}_{l_M}} \right\},\label{eq_Phi_r_l}
\end{equation}
where $\beta^{F}_{l_m}$ and $\beta^{B}_{l_m}$ are amplitudes of the forward and backward sides for STAR-RIS $l$'s element $m$. $\theta^{F}_{l_m}$ and $\theta^{B}_{l_m}$ are phase shift coefficients of the forward and backward sides for STAR-RIS $l$'s element $m$. The channel gain $h_{b, u}$ from AP $b$ to UE $u$ can be formulated as:
\begin{equation}
h_{b,u}  = \sqrt{\frac{\kappa}{\kappa+1}}h^{\textrm{LoS}}_{b,u} + \sqrt{\frac{1}{\kappa+1}}h^{\textrm{NLoS}}_{b,u},\label{eq_h_b_u}
\end{equation}
where $\kappa$ is the Rician factor, $h^{\textrm{LoS}}_{b,u} \in \mathbb{C}^{N_{b} \times 1}$ and $h^{\textrm{NLoS}}_{b,u} \in \mathbb{C}^{N_{b} \times 1}$ are LoS and NLoS channel gains between AP $b$ and UE $u$, where $\mathbb{C}^{N_{b} \times 1}$ denotes a complex matrix of size $N_{b} \times 1$. Similarly, we can define the channel gain $g_{b, l}$ from AP $b$ to the STAR-RIS $l$ and the channel gain $g_{l, u}$ from STAR-RIS $l$ to UE $u$ as follows:
\begin{equation}
g_{b,l} = \sqrt{\frac{\kappa}{\kappa+1}}g^{\textrm{LoS}}_{b,l} + \sqrt{\frac{1}{\kappa+1}}g^{\textrm{NLoS}}_{b,l},\label{eq_g_b_l}
\end{equation}
\begin{equation}
g_{l,u} = \sqrt{\frac{\kappa}{\kappa+1}}g^{\textrm{LoS}}_{l,u} + \sqrt{\frac{1}{\kappa+1}}g^{\textrm{NLoS}}_{l,u},\label{eq_g_l_u}
\end{equation}
where $g^{\textrm{LoS}}_{b,l} \in \mathbb{C}^{N_{b} \times M}$ and $g^{\textrm{NLoS}}_{b,l} \in \mathbb{C}^{N_{b} \times M}$ \textcolor{black}{are LoS and NLoS channel gains from AP $b$ to STAR-RIS $l$, and $g^{\textrm{LoS}}_{l,u} \in \mathbb{C}^{M \times 1}$ and $g^{\textrm{NLoS}}_{l,u} \in \mathbb{C}^{M \times 1}$ are LoS and NLoS channel gains from STAR-RIS $l$ to UE $u$.} Hence, the combined channel gain from AP $b$ to UE $u$ is given by 
\begin{equation}
\hat{h}_{b,u} = 
\begin{cases}
h_{b,u} + \sum_{l\in\mathcal{L}} \left\{ c^{b}_{l}(c^{l_{F}}_{u}g_{b,l} \boldsymbol{\Phi}^{F}_{l} g_{l,u} + c^{l_{B}}_{u}g_{b,l} \boldsymbol{\Phi}^{B}_{l} g_{l,u}) \right\} \\ \hfill \text{if} \ \ {c^{b}_{u}=1}, \\
\sum_{l\in\mathcal{L}} \left\{  c^{b}_{l}(c^{l_{F}}_{u}g_{b,l} \boldsymbol{\Phi}^{F}_{l} g_{l,u} + c^{l_{B}}_{u}g_{b,l} \boldsymbol{\Phi}^{B}_{l} g_{l,u})  \right\} \\ \hfill \text{if} \ \ {c^{b}_{u}=0},
\end{cases} \label{eq_hat_h_b_u}
\end{equation}
\textcolor{black}{where $c^{b}_{u},c^{b}_{l},c^{l_{F}}_{u},c^{l_{B}}_{u} \in \left\{ 0,1 \right\}$ are adjacency indicators between AP $b$ and UE $u$, between AP $b$ and STAR-RIS $l$, between a forward side of STAR-RIS $l_{F}$ and UE $u$, and between a backward side of STAR-RIS $l_{B}$ and UE $u$.}

\textcolor{black}{In NOMA networks, intra-cluster and inter-cluster interference can be considered, where intra-cluster interference occurs between UEs grouped in the same cluster of each AP, and inter-cluster interference occurs between UEs grouped in different clusters of each AP. Thus, we first introduce the user association indicator as $\alpha_{b,u} \in \left\{ 0,1 \right\}$, where $\alpha_{b,u}=1$ if UE $u$ is associated with AP $b$, and $0$ otherwise. Then, the UEs associated with AP $b \in \mathcal{B}$ are further clustered into $K_b$ groups. Therefore, we define $\gamma^{b}_{k,u} \in \left\{ 0,1 \right\}$ as a user pairing factor, where $\gamma^{b}_{k,u}=1$ if UE $u$ is involved in the cluster $k$ of AP $b$, otherwise $\gamma^{b}_{k,u}=0$.} \textcolor{black}{ Moreover, let $\omega_{b} = \left\{ w_{b,1}, w_{b,2}, ... , w_{b,K_{b}} \right\}$ be the active beamforming vector of AP $b$.} Therefore, the received signal of UE $u$ associated with AP $b$ in cluster $k$ can be given by
\begin{multline}
y^{b}_{k,u} = \hat{h}_{b,u} \bigg[\omega_{b,k}\left(\alpha_{b,u}\gamma^{b}_{k,u}p_{b}s^{b}_{k,u} + \sum^{U}_{u' \ne u} \alpha_{b,u'}\gamma^{b}_{k,u'}p_{b}s^{b}_{k,u'}\right) \\ + \sum^{K_{b}}_{k' \ne k} \sum^{U}_{u'} \alpha_{b,u'}\gamma^{b}_{k',u'}\omega_{b,k'}p_{b}s^{b}_{k',u'}\bigg] + N_{0},\label{eq_y_b_u}
\end{multline}
where \textcolor{black}{$p_{b}$ is power allocation coefficient of each UE associated with AP $b$. We assume that UEs connected to the same AP use power equally. Therefore, the power allocation coefficient for AP $b$ satisfies $p_{b}=1/\sum^{\mathcal{U}}\alpha_{b,u}$.} $s^{b}_{k,u}$ denotes the signal transmitted by AP $b$ for UE $u$ in cluster $k$, and $N_{0}$ is the additive white Gaussian noise (AWGN) with variance $\sigma^{2}$. Without loss of generality, for any cluster $k \in \mathcal{K}_{b}$, $\delta_{b,k}(u)$ denotes the UE index that corresponds to UE $u$ decoded order in the SIC procedure. For cluster $k$, after applying the SIC decoding procedure \cite{cui2018unsupervised}, the intra-cluster and inter-cluster powers of UE $u$ associated with AP $b$ on cluster $k$ can be given by:
\color{black}
\begin{equation}
I^{b,\textrm{intra}}_{k,u} =|\hat{h}_{b,u}\omega_{b,k}|^{2}\sum^{U}_{\delta_{b,k}(u') > \delta_{b,k}(u)} \alpha_{b,u'}\gamma^{b}_{k,u'}p_{b}, \label{eq_intra}
\end{equation}
\begin{equation}
I^{b,\textrm{inter}}_{k,u} = \sum^{K_{b}}_{k' \ne k} \sum^{U}_{u'} \alpha_{b,u'}\gamma^{b}_{k',u'}|\hat{h}_{b,u}\omega_{b,k'}|^{2} , \label{eq_inter}
\end{equation}
Accordingly, the received signal-to-interference-plus-noise ratio (SINR) of UE $u$ associated with AP $b$ in cluster $k$ is:
\begin{equation}
\textrm{SINR}^{b}_{k,u} = {\frac{|\hat{h}_{b,u}\omega_{b,k}|^{2}\alpha_{b,u}\gamma^{b}_{k,u}p_{b}}{I^{b,\textrm{intra}}_{k,u} + I^{b,\textrm{inter}}_{k,u} + \sigma^2}}, \label{eq_sinr}
\end{equation}
For any two UEs $v$ and $u$ with decoding order $\delta_{b,k}(v) > \delta_{b,k}(u)$ in the same AP $b$ and cluster $k$, the received SINR of the signal $s^{b}_{k,u}$ at the UE $v$ is given by:
\begin{equation}
\textrm{SINR}^{b}_{k,v \rightarrow u} = {\frac{|\hat{h}_{b,v}\omega_{b,k}|^{2}\alpha_{b,v}\gamma^{b}_{k,v}p_{b}}{I^{b,\textrm{intra}}_{k,v \rightarrow u} + I^{b,\textrm{inter}}_{k,v \rightarrow u} + \sigma^2}}, \label{eq_sinr_v_u}
\end{equation}
where $I^{b,\textrm{intra}}_{k,v \rightarrow u}=|\hat{h}_{b,v}\omega_{b,k}|^{2}\sum^{U}_{\delta_{b,k}(u') > \delta_{b,k}(u)} \alpha_{b,u'}\gamma^{b}_{k,u'}p_{b}$ is the intra-cluster interference power of the signal $s^{b}_{k,u}$ at UE $v$. $I^{b,\textrm{inter}}_{k,v \rightarrow u} = \sum^{K_{b}}_{k' \ne k} \sum^{U}_{u'} \alpha_{b,u'}\gamma^{b}_{k',u'}|\hat{h}_{b,v}\omega_{b,k'}|^{2}$ is the inter-cluster interference power of the signal $s^{b}_{k,u}$ at UE $v$. It is worth pointing out that given a decoding order, to guarantee the SIC performed successfully, the condition $\textrm{SINR}^{b}_{k,v \rightarrow u} \geq \textrm{SINR}^{b}_{k,u}$ with $\delta_{b,k}(v) > \delta_{b,k}(u)$ must be guaranteed. Therefore, the achievable data rate of UE $u$ associated with AP $b$ in cluster $k$ is calculated as:
\begin{equation}
R^{b}_{k,u} = \log_2 \left( 1 + \textrm{SINR}^{b}_{k,u} \right), \label{eq_R_b_u}
\end{equation}

\subsection{Problem Formulation}
\label{prob_form}
We now define the detailed problem formulation based on the proposed system model. This network's primary goal is to maximize the achievable sum rate of $U$ UEs (considered as a network utility), while jointly optimizing user association $\boldsymbol{\alpha}$, user pairing factor $\boldsymbol{\gamma}$, decoding order $\boldsymbol{\delta}$, active beamforming $\boldsymbol{\omega}$, and passive beamforming $\boldsymbol{\Phi}=\left\{\Phi^{F}, \Phi^{B}\right\}$ of Multi-STAR-RIS. Therefore, we can define our optimization problem as follows:
\begin{maxi!}[2]<b> 
    {\substack{\boldsymbol{\alpha}, \boldsymbol{\gamma}, \boldsymbol{\delta}, \boldsymbol{\omega}, \boldsymbol{\Phi}}}   
    {\sum_{b=1}^{B} \sum_{k=1}^{K_{b}} \sum_{u=1}^{U}  R^{b}_{k,u}}{\label{opt:P1}}{\textbf{P1:}}
    \addConstraint{\sum_{b=1}^{B} \sum_{k=1}^{K_{b}} R^{b}_{k,u} \geq R^{\textrm{min}}_u,~\forall u \in \mathcal{U} \label{P1_C1}}
    \addConstraint{
    \begin{aligned}
    \textrm{SINR}^{b}_{k,v \rightarrow u} \geq & \textrm{SINR}^{b}_{k,u},
    \\&\delta_{b,k}(v) > \delta_{b,k}(u), 
    \end{aligned}
    \label{P1_C9}}
    \addConstraint{
    \begin{aligned}
          \alpha_{b,u}, \gamma^{b}_{k,u} \in & \left\{0,1\right\}, \\& ~\forall b \in \mathcal{B}, ~\forall k \in \mathcal{K}_{b},~\forall u \in \mathcal{U},{\label{P1_C2}}
    \end{aligned}
    }
    \addConstraint{ \sum_{b=1}^{B} \alpha_{b,u}=1,~\forall u \in \mathcal{U} \label{P1_C3}}
    \addConstraint{ \sum_{u=1}^{U} \alpha_{b,u} \leq Q^{b},~\forall b \in \mathcal{B} \label{P1_C10}}
    \addConstraint{ \sum_{k=1}^{K_{b}} \lVert w_{b,k} \rVert^{2} \leq P_{\textrm{max}},~\forall b \in \mathcal{B} \label{P1_C5}}
    \addConstraint{ \sqrt{\beta^{t}_{l_m}}, \sqrt{\beta^{e}_{l_m}} \in [0,1],~\forall l \in \mathcal{L},~\forall m \in \mathcal{M} \label{P1_C6}}
    \addConstraint{ \beta^{t}_{l_m} + \beta^{r}_{l_m} = 1,~\forall l \in \mathcal{L},~\forall m \in \mathcal{M} \label{P1_C7}}
    \addConstraint{ \theta^{t}_{l_m},\theta^{r}_{l_m}\in [0,2\pi),~\forall l \in \mathcal{L},~\forall m \in \mathcal{M}, \label{P1_C8}}
\end{maxi!}
\textcolor{black}{where $R^{\textrm{min}}_u$ is the minimum rate requirement of each UE. Constraint (\ref{P1_C1}) guarantees the QoS requirement of each UE, and constraint (\ref{P1_C9}) ensures the success of the SIC decoding. Furthermore, constraint (\ref{P1_C2}) represents the binary variables. Constraints (\ref{P1_C3}) and (\ref{P1_C10}) guarantee that each UE can be associated with only one AP and each AP $b$ serves at most $Q^{b}$ UEs in the network. Constraint (\ref{P1_C5}) ensures the power budget constraint of each AP. Finally, Constraints (\ref{P1_C6}) to (\ref{P1_C8}) indicate the requirements of each reflecting and transmission element in STAR-RIS. To solve this proposed problem, we provide a solution approach in the next section.}
 
\section{Joint Optimization of Multi-STAR-RIS}
\label{sol_app}
\begin{figure}[t]
    \centering
    \includegraphics[width=\columnwidth]{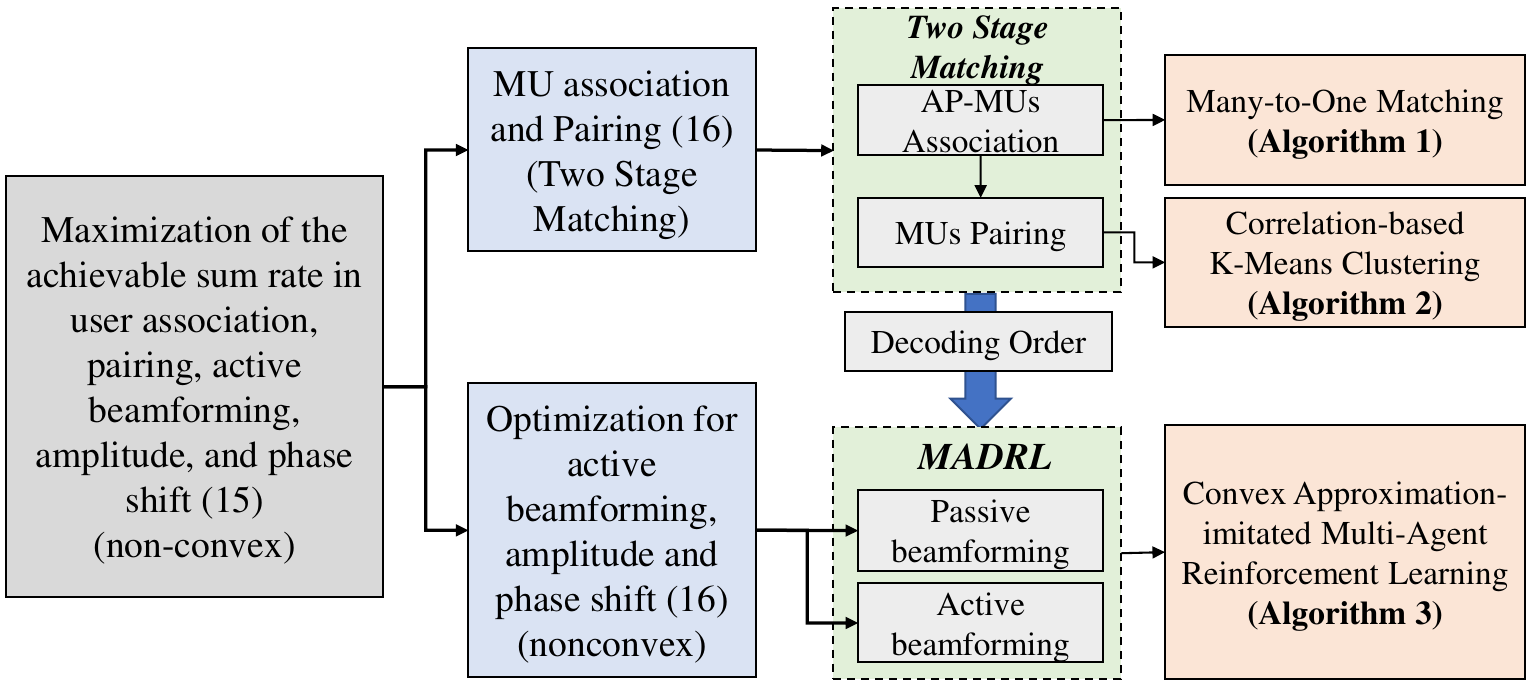}
    \caption{The proposed decomposition and solution framework for each subproblem and associated algorithm.}
    \label{fig_roadmap}
\end{figure}

\textcolor{black}{The transmission and reflection coefficients by Multi-STAR-RIS are highly linked in comparison to the conventional NOMA network or the conventional RIS that uses only reflection coefficients. Thus,} it can be observed that the formulated problem in (\ref{opt:P1}) is a mixed-integer nonlinear programming (MINLP) problem, which is a non-polynomial-time (NP-hard) problem. Therefore, getting an optimal solution within polynomial time is impossible. Thus, to solve the problem realistically, we decompose the main problem into two sub-problems: 1) UE assignment and pairing problem, and \textcolor{black}{2) beamforming vectors optimization for APs and Multi-STAR-RIS.} Then, the decomposed sub-problems are solved, sequentially. \textcolor{black}{Fig.~\ref{fig_roadmap} presents an overview and workflow of the proposed problem decomposition and approach for each sub-problem.}

\subsection{Two-Stage Matching for UE Association and Pairing}
Before the active and passive beamforming optimization, we try to reduce the complexity of the problem (\ref{opt:P1}) by separating AP-UE Association and UE Pairing. Therefore, given the active and passive beamforming vector, we first try to solve the 2-stage Matching problem for UE Association and Pairing, which can be rewritten as
\label{sol_3dmatching}
\begin{maxi!}[2] 
		{\substack{\boldsymbol{\alpha}, \boldsymbol{\gamma}}}   
		{\sum_{b=1}^{B} \sum_{k=1}^{K_{b}} \sum_{u=1}^{U}  R^{b}_{k,u}}{\label{opt:P1.1}}{\textbf{P1.1:}}
		\addConstraint{\textcolor{black}{(\ref{P1_C1})\sim({\ref{P1_C10}}).}}
\end{maxi!}
We propose a many-to-one matching for UE association and correlation-based K-means clustering for solving, respectively.
\subsubsection{Many-to-One Matching-based AP-UE Association}
\label{sol_m2o}
The UE association problem can be modeled as a many-to-one matching game \cite{ren2021matching}. The concept of many-to-one matching refers to a scenario where agents belonging to one group can be matched with multiple agents from another group. In this game, there are two disjoint sets of players: the set of UE clusters $\mathcal{K}$, and the set of APs $\mathcal{B}$. In our proposed matching game, each UE cluster $k \in \mathcal{K}$ can be associated with at most one AP. Moreover, each AP can serve a certain number of clusters depending on the maximum number of allowable clusters $Q^b$ at AP $b \in \mathcal{B}$. Assume that each UE cluster $k$ has a preference list $P(k)=b_{3}, b_{2}, b_{1}, b_{4}, ...$ which is a list in which APs are sorted in order of UE cluster $k$'s preference. For example, UE cluster $k$ prefer AP $b_{3}$ to AP $b_{2}$. Therefore, many-to-one matching can be defined as\\
\textbf{Definition 1.} \textit{A matching $\mu$ is a function from the set $\mathcal{B} \cup \mathcal{K}$ into the set of unordered families of elements of $\mathcal{B} \cup \mathcal{K}$ such that:\\
(1) $|\mu(k)|=1$ for each $k \in \mathcal{K}$ and $\mu(k)=k$ if $u$ is unassociated;\\
(2) $|\mu(b)| \leq Q^{b}$ for each AP $b \in \mathcal{B}$;\\
(3) $\mu(k)=b$ if and only if $k$ is associated with $b$ and is an element of $\mu(b)$.}\\
where $|\mu(b)|=l$ means that $l$ UE clusters are associated with AP $b$ and $\mu(b)=\left\{ k_{1}, k_{2},...,k_{l} \right\}$ is a set of UE clusters associated with AP $b$.

For maximizing the achievable rate of UEs, each AP $b$ determines the achievable rate with each UE cluster $k$ and has a preference list in high order. In the many-to-one matching problem of UE association, we define the preference of each UE cluster $k$ associated with AP $b$ as
\begin{equation}
U_{b,k} = \sum_{u \in C_{k}} R^{b}_{k,u}. \label{eq_pre_cluster}
\end{equation}
Then we define the preference of each AP $b$ as
\begin{equation}
U_{b} = \sum_{k \in \mathcal{K}}\sum_{u \in C_{k}} R^{b}_{k,u}. \label{eq_pre_ap}
\end{equation}
Thus, in this matching, each AP $b$ has a strict preference ordering $\succ_{b}$ over $\mathcal{K}$. Each UE cluster also has a preference relation $\succ_{k}$ over the set $\mathcal{B} \cup \left\{0\right\}$, where $\left\{0\right\}$ denotes the UE cluster is unmatched. Specifically, for a given UE cluster $k$, any two APs $b$ and $b'$ with $b,b' \in \mathcal{B}$, any two matchings $\mu$ and $\mu'$ are defined as
\begin{equation}
(b,\mu) \succ_k (b',\mu') \Leftrightarrow U_{b,k}(\mu) > U_{b,k}(\mu'),
\end{equation}
which indicates that the UE cluser $k$ prefers AP $b$ in $\mu$ to BS $b'$ in $\mu'$ only if the UE cluser $k$ can achieve a higher rate on AP $b$ than AP $b'$. Analogously, for any AP $b$, its preference over the UE cluster set can be described as follows. For any two subsets of UE clusters $\mathcal{S}$ and $\mathcal{S}'$ with $\mathcal{S} \neq \mathcal{S}'$, any two matchings $\mu$ and $\mu'$ with $\mathcal{S} = \mu(b)$ and $\mathcal{S}' = \mu'(b)$ are defined as
\begin{equation}
(\mathcal{S},\mu) \succ_b (\mathcal{S}',\mu') \Leftrightarrow U_{b}(\mu) > U_{b}(\mu'),
\end{equation}
which represnets that AP $b$ prefers the set of UE clusters $\mathcal{S}$ to $\mathcal{S}'$ only when AP $b$ can get a higher rate from $\mathcal{S}$. Due to the existence of peer effects and non-substitutability in (\ref{eq_sinr}), the preference lists of players change frequently during the matching process, which makes it difficult to design the matching processes. To handle the peer effects and ensure exchange stability, Given a matching function $\mu$, and assume that $\mu(i)=n$ and $\mu(j)=m$, we define the swap matching as
\begin{equation}
\mu^{j}_{i}=\left\{ \mu \setminus \left\{(i,n),(j,m)\right\} \bigcup \left\{(j,n),(i,m)\right\} \right\},
\label{eq_swap_matching}
\end{equation}
where UE clusters $i$ and $j'$ exchange their matched APs $n$ and $m$ while keeping all other matching states the same. Based on the swap operation in (\ref{eq_swap_matching}), we define the concept of swap-blocking pair as follows \cite{cui2017optimal}.\\
\textbf{Definition 2.} \textit{A matching $\mu$ is two-sided exchange-stable if and only if there does not exist a pair of UE clusters $(i,j)$ such that:\\
(1) $\forall k \in \left\{i,j,n,m\right\},U_{k}(\mu^{j}_{i}) \geq U_{k}(\mu)$;\\
(2) $\exists k \in \left\{i,j,n,m\right\},$ such that $U_{k}(\mu^{j}_{i}) > U_{k}(\mu)$,}\\
where $U_{k}(\mu)$ represents the utility of UE cluster $k$ under matching $\mu$.
However, swap matching depends on the performance given the initial matching. Therefore, to achieve the goal of stable initial matching, The deferred acceptance (DA) algorithm was proposed and applied to the marriage markets and college admission problems \cite{gale1962college}. In the DA algorithm, the agents on one group propose a pair formation with the agents of the other side group according to their preference, and an iterative procedure. At first, the matching procedure starts with building preference lists, i.e., $P(k)$ for UE cluster $k$. In each iteration, each UE cluster $k$ proposes to highest preferred AP for association in the preference list. Each AP $b$ accepts the proposal of UE cluster $k^{*}$ with the highest rank among the UE cluster's proposal list $G(b)$. At the same time, excludes AP $b$ from UE cluster $k^{*}$'s preferred list. However, when the current number of proposals for AP $b$ is greater than the maximum capacity, i.e., $|\mu(b)| \geq Q^{b}$, if a proposal from UE $k^{*}$ comes, no further proposals are accepted and AP $b$ is excluded from the preference list of UE cluster $k^{*}$. The iterative approach continues until every UE cluster has gotten an acceptable proposal, at which point a stable solution to the UE association problem is attained. Finally, the output of the many-to-one matching $\mu$ is transformed to the UE association vector $\boldsymbol{\alpha}$. The complexity of a many-to-one matching game-based DA algorithm depends on the required number of accepting/rejecting decisions to attain the stable matching $\mu$. Each UE cluster in the network recommends an association with the AP at the top of their preference list during each iteration. The AP then decides whether to accept or reject the proposal. Each UE cluster's preference list can only be as large as $|\mathcal{B}|$ in this case. As a result, the stable matching converges in $\mathcal{O}(\mathcal{U}\mathcal{B})$ iterations, where $\mathcal{U}$ and $ \mathcal{B}$ are the number of UEs and APs in the considered network. Algorithm \ref{alg:association} describes the many-to-one matching based on the DA algorithm for UE association. \textcolor{black}{When the UE's location changes or the channel state changes, the stable matching may change due to the peer effect. Thus, when the network state changes, a new matching may be searched again through Algorithm \ref{alg:association} to find a stable matching suitable for the changed environment.}
\begin{algorithm}[t]
	\caption{\strut Many-to-One Matching for UE Association} 
	\label{alg:association}
	\begin{algorithmic}[1]
	    \STATE{\textbf{Step-I:} Initial matching based on DA algorithm}
        \STATE{The sorted preference lists $P(k)$ for all UE clusters based on the scalar channel gain $\sum_{u \in C_{k}}|\hat{h}_{b,u}|^{2}$.}
        \STATE{$\mathcal{K}_{\textrm{cur}} \leftarrow \mathcal{K}$}
        \WHILE{$\mathcal{K}_{\textrm{cur}} \neq \emptyset$}
        \STATE{$G(b) \leftarrow \emptyset, \forall b \in \mathcal{B}$}
        \FOR{$k \in \mathcal{K}_{\textrm{cur}}$}
        \STATE{$G(b)=G(b) \cup \left\{k\right\}$, with $P(k)[0]=b$}
        \ENDFOR        
        \FOR{$b \in \mathcal{B}$}
        \IF{$|G(b)| \neq 0$}
        \STATE{Sort $G(b)$ by the preference with the scalar channel gain $\sum_{u \in C_{k}}|\hat{h}_{b,u}|^{2}$}
        \STATE{$k^{*} = G(b)[0]$}
        \IF{$|\mu(b)|<Q^{b}$}
        \STATE{$\mu(b)=\mu(b)\cup \left\{k^{*}\right\}$}
        \STATE{$\mu(k^{*})=b$}
        \STATE{$\mathcal{K}_{\textrm{cur}} = \mathcal{K}_{\textrm{cur}} \setminus \left\{k^{*}\right\}$}
        \ELSE
        \STATE{$P(k^{*})=P(k^{*}) \setminus \left\{b\right\} $}
        \ENDIF
        \ENDIF
        \ENDFOR        
        \ENDWHILE
		\STATE{\textbf{Output:} The initial matching function $\mu^{0}$.}
	    \STATE{\textbf{Step-II:} Two-sided exchange stable matching for UE Association}
        \STATE{The initial matching function $\mu^{0}$.}
        \REPEAT
        \STATE{$\forall$ UE cluster $k \in \mu^{0}$, it searches for another UE cluster $k' \in \mu^{0} \setminus \mu^{0}(\mu(k))$ to check whether $(k,k')$ is a swap-blocking pair.}
        \IF{$(k,k')$ is a swap-blocking pair}
        \STATE{Update $\mu = \mu^{k'}_{k}$}
        \ELSE
        \STATE{Keep the current matching state}
        \ENDIF
        \UNTIL{No swap-blocking pair can be constructed.}
		\STATE{\textbf{Output:} The optimal matching function $\mu^{*} \rightarrow$ UE association vector $\boldsymbol{\alpha}^*$.}
	\end{algorithmic}
\end{algorithm}

\subsubsection{Correlation-Based K-Means Clustering for UE Pairing}
\label{sol_pairing}
\begin{algorithm}[t]
	\caption{\strut Correlation-Based K-Means Clustering for UE Pairing} 
	\label{alg:pairing}
	\begin{algorithmic}[1]
	    \STATE{\textbf{Input:} the initial passive beamforming vector $\boldsymbol{\Phi}_0$, the given user assignment $\boldsymbol{\alpha^{*}}$.}
        \IF{$b \in \mathcal{B}$}
        \STATE{$C^{b}_{\textrm{pre}}, C^{b} = \emptyset$}
        \STATE{Randomly select $\chi^{b}_{k}$ from $u \in \mathcal{U}_{b}$ , $\forall k=1,...,K$}
        \WHILE{$C^{b}_{\textrm{pre}} \neq C^{b}$}
        \STATE{$C^{b}_{\textrm{pre}} \leftarrow C^{b}$}
        \FOR{$u \in \mathcal{U}_{b}$ with $u \neq \chi^{b}_{k}$ , $\forall k=1,...,K_{b}$}
        \STATE{$k^{*}=\arg\max_{1 \leq k^{*} \leq K_{b}}\textrm{Cor}^{b}_{u,\chi_{k^{*}}}$}
        \STATE{$C^{b}_{k^{*}}=C^{b}_{k^{*}} \cup \left\{ u \right\}$}
        \ENDFOR
        \STATE{Update $\chi^{b}_{k}$ according to (\ref{eq_representative}), $\forall k=1,...,K_{b}$}
        \STATE{$C^{b}_{k}$ = $C^{b}_{k} \setminus u$, $\forall u \neq \chi^{b}_{k} \in C^{b}_{k}$, $\forall k=1,...,K_{b}$}
        \ENDWHILE
        \ENDIF
		\STATE{\textbf{Output:} The optimal clustering vector $\mathbf{C}^{*} \rightarrow$ UE pairing vector $\boldsymbol{\gamma}^*$.}
	\end{algorithmic}
\end{algorithm}

UEs whose channels are highly correlated should be assigned to the same group to make full use of the multiplexing gain, while UEs whose channels are uncorrelated should be assigned to different groups to decrease interference. We adopt the K-means clustering algorithm to implement the UE pairing in the NOMA network. K-means clustering is one way to divide given data into multiple partitions \cite{hartigan1979algorithm}. The K-means algorithm determines the cost function as the sum of squares at the center of each group and the group's distance from the data subject. Clustering is also performed by updating the group that each data object belongs to to minimize the value of this cost function. Therefore, we use the channel correlation between each UE for the cost function for UE pairing. The normalized channel correlation between UE $i$ and UE $j$ in the same AP $b$ can be calculated as \cite{zhu2019millimeter}:
\begin{equation}
\textrm{Cor}^{b}_{i,j} = {\frac{\hat{h}^{\textrm{H}}_{b,i}\hat{h}_{b,j}}{\lVert \hat{h}_{b,i} \rVert \lVert \hat{h}_{b,j} \rVert}}. \label{eq_correlation}
\end{equation}
In order to cluster UEs connected to each AP $b$ into $K_b$, we select randomly $K_b$ UEs assigned to clusters, $C^{b} = \left\{C^{b}_{1}, C^{b}_{2},..., C^{b}_{K_{b}}\right\}$, one by one. Then, the channel correlation between the unselected UE $u'$ and the selected UE $u$ is calculated based on (\ref{eq_correlation}), and the UE $u'$ having the highest channel correlation is assigned to the cluster to which the UE $u$ belongs. Therefore, each representative can be selected from each cluster. The representative of each cluster is updated as the one with the lowest correlation with the other clusters to further reduce the correlation of the channels between the various clusters. The correlation between a UE to the other clusters is the total normalized channel correlation between an UE to the UEs of the other clusters. The correlation between a UE $u$ in the cluster $C^{b}_k$ at the AP $b$ to the other clusters is:
\begin{equation}
\Bar{\textrm{Cor}}^{b}_{u} = \sum^{K}_{l \neq k} \sum_{u' \in C_{l}} \textrm{Cor}^{b}_{u,u'}. \label{eq_correlation_cluster}
\end{equation}
After that, the representative $\chi^{b}_{k}$ of the cluster $C^{b}_k$ is updated as:
\begin{equation}
\chi^{b}_{k} = \arg\min_{u \in C^{b}_{k}} \Bar{\textrm{Cor}}^{b}_{u}. \label{eq_representative}
\end{equation}
Following the update of the representative for each cluster, the other UEs are subsequently reassigned to their respective clusters. The iteration is terminated when the representatives of the clusters remain unaltered. Finally, the optimal output $\mathbf{C}^{*}$ of the clustering vector, is transformed to the UE pairing vector $\boldsymbol{\gamma}^{*}$. The Correlation-based K-means Clustering is described in Algorithm \ref{alg:pairing}. \textcolor{black}{When executing the UE paring in Algorithm \ref{alg:pairing}, the computational complexity for calculating the channel correlation is $\mathcal{O}(\mathcal{K}^{2}\mathcal{U})$, where $\mathcal{K}$ the number of cluster of AP. During each iteration, the computational complexity for updating the cluster representative and the user grouping are $\mathcal{O}(\mathcal{K}^{2})$ and $\mathcal{O}(\mathcal{K}\mathcal{M})$, where $\mathcal{M}$ the number of RF chain, respectively. Therefore, the UE pairing computational complexity for all APs belongs to $\mathcal{O}(\mathcal{B}\mathcal{N}(\mathcal{K}^{2}\mathcal{U}+\mathcal{K}^{2}+\mathcal{K}\mathcal{M}))$.}

\subsection{Decoding Order}
\label{sol_decoding}
Before handling the beamforming optimization problems, the decoding order must be addressed because it is an important one for the Multi-STAR-RIS in the NOMA network. Therefore, we propose a scheme to obtain the optimal decoding order by the following lemma.\\
\textbf{Lemma 1.}~\textit{Given the active beamforming vector $\boldsymbol{\omega}$ and the passive beamforming vector $\boldsymbol{\Phi}$, The decoding order for cluster $k$ with $|C^{b}_{k}|$ UEs in AP $b$ is defined as \cite{cui2018unsupervised}, \cite{zuo2022joint}.
\begin{equation}
g^{b,k}_{\delta_{b,k(1)}} \leq g^{b,k}_{\delta_{b,k(2)}} \leq \cdots \leq g^{b,k}_{\delta_{b,k}(|C^{b}_{k}|)}, \label{eq_decoding_order}
\end{equation}
where $g^{b,k}_{\delta_{b,k(j)}}={\frac{|\hat{h}_{b,u}\omega_{b,k}|^{2}}{\sum^{K_{b}}_{k' \ne k} \sum_{u'\in C^{b}_{k'}}|\hat{h}_{b,u}\omega_{b,k'}|^{2} + \sigma^2}}$  is the equivalent-combined channel gain \cite{zuo2022joint, higuchi2013non}.}

Lemma 1 indicates that the decoding order for each cluster of the Multi-STAR-RIS in the NOMA system is a function of the active beamforming vectors $\boldsymbol{\omega}$, the passive beamforming vectors $\boldsymbol{\Phi}$. For any two users $u$ and $v$ belong to cluster $k$ in the AP $b$, if the decoding order of the two users satisfies
\begin{equation}
\delta_{b,k}^{-1}(v) > \delta_{b,k}^{-1}(u),
\end{equation}
where $\delta_{b,k}^{-1}(\cdot)$ is the inverse of mapping function $\delta_{b,k}(\cdot)$. Then, under the optimal decoding order, the following SIC condition is guaranteed:
\begin{equation}
\textrm{SINR}^{b}_{k,v \rightarrow u} \geq \textrm{SINR}^{b}_{k,u}.
\end{equation}
\begin{IEEEproof}
For any two users $u$ and $v$ belonging to cluster $k$ in the AP $b$ with the optimal decoding order $\delta_{b,k}^{-1}(v) > \delta_{b,k}^{-1}(u)$, according to Lemma 1, the equivalent combined channel gains of the two users have to satisfy the following condition
\begin{multline}
    {\frac{|\hat{h}_{b,v}\omega_{b,k}|^{2}}{\sum^{K_{b}}_{k' \ne k} \sum_{v'\in C^{b}_{k'}}|\hat{h}_{b,v}\omega_{b,k'}|^{2} + \sigma^2}} \geq \\ {\frac{|\hat{h}_{b,u}\omega_{b,k}|^{2}}{\sum^{K_{b}}_{k' \ne k} \sum_{u'\in C^{b}_{k'}}|\hat{h}_{b,u}\omega_{b,k'}|^{2} + \sigma^2}} \label{A_1}
\end{multline}
(\ref{A_1}) can be reformulated as
\begin{multline}
{|\hat{h}_{b,v}\omega_{b,k}|^{2}} \left( {\sum^{K_{b}}_{k' \ne k} \sum_{u'\in C^{b}_{k'}}|\hat{h}_{b,u}\omega_{b,k'}|^{2} + \sigma^2} \right) \\ \geq {|\hat{h}_{b,u}\omega_{b,k}|^{2}} \left( {\sum^{K_{b}}_{k' \ne k} \sum_{v'\in C^{b}_{k'}}|\hat{h}_{b,v}\omega_{b,k'}|^{2} + \sigma^2}\right) \label{A_2}
\end{multline}
After multiplying both sides of (\ref{A_2}) by $p_b$, we proceed by adding $p_{b}|\hat{h}_{b,v}\omega_{b,k}|^{2}|\hat{h}_{b,u}\omega_{b,k}|^{2}\sum^{U}_{\delta_{b,k}(u') > \delta_{b,k}(u)} p_{b}$ to both sides. The inequality can be derived as follow
\begin{multline}
|\hat{h}_{b,v}\omega_{b,k}|^{2}p_b \left( {I^{b,\textrm{intra}}_{k,u} + I^{b,\textrm{inter}}_{k,u} + \sigma^2} \right) \\ \geq |\hat{h}_{b,u}\omega_{b,k}|^{2}p_b \left( {I^{b,\textrm{intra}}_{k,v \rightarrow u} + I^{b,\textrm{inter}}_{k,v \rightarrow u} + \sigma^2} \right).
\label{A_3}\end{multline}
Rearranging $(\ref{A_3})$, we have the following:
\begin{equation}
\textrm{SINR}^{b}_{k,v \rightarrow u} \geq \textrm{SINR}^{b}_{k,u}, \label{A_4}
\end{equation}
which means that the SIC condition $R^{b}_{k,v \rightarrow u} \geq R^{b}_{k,u}$.
\end{IEEEproof} 

According to Proposition 1, the constraint in (\ref{P1_C9}) can be removed under the optimal decoding order of the NOMA system. This operation will not affect the optimality of the problem (\ref{opt:P1}). Furthermore, Lemma 1 guarantees that once the association, pairing, and beamforming vectors are determined, the optimal decoding order in each cluster is fixed \cite{zuo2022joint, cui2018unsupervised}. Thus, we developed the optimal beamforming vectors for UEs in the Multi-STAR-RIS NOMA based on this observation.

\subsection{Convex Approximation imitated MARL Scheme}
Since there is no inter-cell interference between APs, the optimization problem in (\ref{opt:P1}) for the beamforming vectors optimization can be decomposed into $B$ decoupled subproblems. At the fixed UE association, pairing, and decoding order, we can rewrite the beamforming vectors optimization problem for any AP $b$ as follows:
\label{sol_rl}
\begin{maxi!}[2]<b> 
		{\substack{\boldsymbol{\omega}, \boldsymbol{\Phi}}}   
		{\sum_{k=1}^{K_{b}} \sum_{u \in C^{b}_{k}}   R^{b}_{k,u}}{\label{opt:P1.2}}{\textbf{P1.2:}}
        \addConstraint{\sum_{k=1}^{K_{b}} R^{b}_{k,u} \geq R^{\textrm{min}}_u,~\forall u \in \mathcal{U} \label{P1.2_C1}}
		\addConstraint{\textcolor{black}{({\ref{P1_C5}})\sim({\ref{P1_C8}}).}}
\end{maxi!}
In this subsection, optimization of active beamforming and passive beamforming through multi-agent reinforcement learning (MARL) is performed. In particular, we first propose a learning scheme that convex approximation imitates reinforcement learning. Existing numerical optimization cannot directly find \emph{global optimal} for non-convex problems, and only partial local optimal can be obtained through conversion to convex problems from non-convex problems. In addition, reinforcement learning can be combined with various communication problems to find the optimal value in a short time through a trained neural model, but as the complexity of the agent's observation, action, and environment increases, convergence speed or convergence cannot be guaranteed. Therefore, we first find a numerical root through a method of iteratively finding a local solution by dividing the active beamforming problem and the passive beamforming problem. In addition, learning proceeds in the direction of reducing the difference between the numerical root and the action selected by the reinforcement learning agent, and gradually reduces the influence on the numerical root so that the reinforcement learning agent can finally derive the global root. As a result, in the next chapter, we first derive the suboptimal value of numerical optimization for imitation learning of the reinforcement learning agent for each active beamforming and passive beamforming and show how to apply the learning scheme to learning through it.

\begin{figure}[t]
    \centering
    \includegraphics[width=\columnwidth]{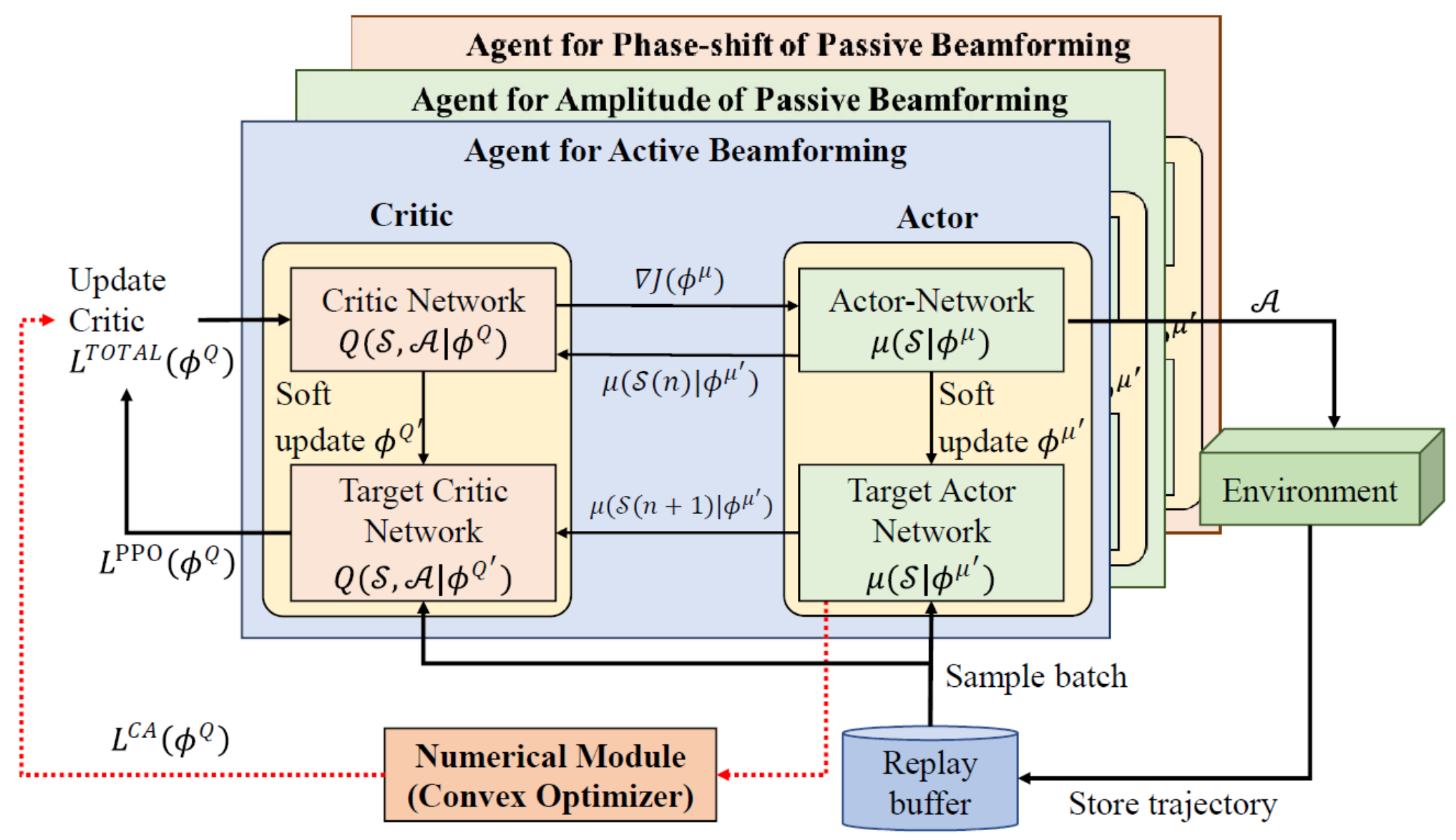}
    \caption{Proposed Convex Approximation-imitated MAPPO.}
    \label{fig_camappo}
\end{figure}

\subsubsection{Active Beamforming based on Convex Approximation}
\label{sol_power}
Given the user association, pairing, and the passive beamforming vector in the problem (\ref{opt:P1}), the active beamforming optimization subproblem for any AP $b$ can be given by:

\begin{maxi!}[2] 
		{\substack{\boldsymbol{\omega}}}   
		{\sum_{k=1}^{K_{b}} \sum_{u \in C^{b}_{k}}  R^{b}_{k,u}}{\label{opt:P1.3}}{\textbf{P1.3:}}
		\addConstraint{\textcolor{black}{(\ref{P1.2_C1})\ \mbox{and} \ ({\ref{P1_C5}}).}}
\end{maxi!}

To transform the subproblem (\ref{opt:P1.3}) into a more tractable form, we introduce slack variables $\left\{ \zeta^{b}_{k,u} |k \in \mathcal{K}_{b}, u \in \mathcal{U}_{b} \right\}$ and $\left\{\eta^{b}_{k,u}|k \in \mathcal{K}_{b}, u \in \mathcal{U}_{b} \right\}$ for any AP $b$ such that
\begin{equation}
\zeta^{b}_{k,u} = {\frac{1}{|\hat{h}_{b,u}\omega_{b,k}|^{2}p_{b}}},\label{eq_zeta}
\end{equation}

\begin{multline}
\eta^{b}_{k,u} = |\hat{h}_{b,u}\omega_{b,k}|^{2}\sum^{U}_{\delta_{b,k}(u') > \delta_{b,k}(u)} p_{b}\\  + \sum^{K_{b}}_{k' \ne k}|\hat{h}_{b,u}\omega_{b,k'}|^{2} + \sigma^2. \label{eq_eta}
\end{multline}

Then, the achievable data rate of UE $u$ associated with AP $b$ in cluster $k$ is calculated as
\begin{equation}
R^{b}_{k,u} = \log_{2}\left(1+{\frac{1}{\zeta^{b}_{k,u}\eta^{b}_{k,u}}}\right) \label{eq_new_datarate}
\end{equation}

Therefore, we can transform the problem (\ref{opt:P1.2}) into
\begin{maxi!}[2]<b> 
		{\substack{\boldsymbol{\omega}, \left\{\zeta^{b}_{k,u}, \eta^{b}_{k,u}, R^{b}_{k,u} \right\} }}   
		{\sum_{k=1}^{K_{b}} \sum_{u \in C^{b}_{k}}  R^{b}_{k,u}}{\label{opt:P1.4}}{\textbf{P1.4:}}
        \addConstraint{\log_{2}(1+{\frac{1}{\zeta^{b}_{k,u}\eta^{b}_{k,u}}}) \geq R^{b}_{k,u},~\forall u \in \mathcal{U} \label{P1.3_C1}}
        \addConstraint{\zeta^{b}_{k,u} \geq \frac{1}{\textrm{Tr}(\mathbf{W}_{b,k}\mathbf{H}_{bku})p_{b}} \label{P1.3_C2}}
        \addConstraint{
        \begin{aligned}
              \eta^{b}_{k,u}  \geq & \textrm{Tr}(\mathbf{W}_{b,k}\mathbf{H}_{bku})  \sum^{U}_{\delta_{b,k}(u') > \delta_{b,k}(u)} p_{b} \\&  + \sum^{K_{b}}_{k' \ne k} \textrm{Tr}(\mathbf{W}_{b,k'}\mathbf{H}_{bku}) + \sigma^2 \label{P1.3_C3}
        \end{aligned}
        }
        \addConstraint{\sum_{k=1}^{K_{b}} \textrm{Tr}(\mathbf{W}_{b,k}) \leq P_{\textrm{max}},\label{P1.3_C4}}
        \addConstraint{\textrm{rank}(\mathbf{W}_{b,k})=1,~\forall k \in \mathcal{K}_{b}\label{P1.3_C54}}
        \addConstraint{\mathbf{W}_{b,k} \succcurlyeq 0,~\forall k \in \mathcal{K}_{b},\label{P1.3_C6}}
\end{maxi!}
where $\mathbf{H}_{bku}=\hat{h}^{H}_{b,u}\hat{h}_{b,u}$ and $\mathbf{W}_{b,k}=\omega^{H}_{b,k}\omega_{b,k}$. Note that $f(x,y)=\log_{2}(1+{\frac{1}{xy}})$ is a convex function for $x>0$ and $y>0$ \cite{mu2020exploiting}. However the subproblem  (\ref{opt:P1.4}) is still con-convex problem due to the constraints (\ref{P1.3_C1}) and (\ref{P1.3_C54}). For the convexity of the constraints (\ref{P1.3_C1}), we apply the first-order Taylor expansion. The lower bound for the RHS of (\ref{P1.3_C1}) with slack variables $\zeta^{b}_{k,u}(n)$ and $\eta^{b}_{k,u}(n)$ can be expressed as
\begin{multline}
\log_{2}\left(1+{\frac{1}{\zeta^{b}_{k,u}\eta^{b}_{k,u}}}\right) \geq \textrm{LOW}^{b}_{k,u} = \\\log_{2}\left(1+{\frac{1}{\zeta^{b}_{k,u}(n)\eta^{b}_{k,u}(n)}}\right) - {\frac{\log_{2}e \left( \zeta^{b}_{k,u} - \zeta^{b}_{k,u}(n) \right)}{\zeta^{b}_{k,u}(n)+{\zeta^{b}_{k,u}(n)}^{2}\eta^{b}_{k,u}(n)}}\\ - {\frac{\log_{2}e \left( \eta^{b}_{k,u} - \eta^{b}_{k,u}(n) \right)}{\eta^{b}_{k,u}(n)+{\eta^{b}_{k,u}(n)}^{2}\zeta^{b}_{k,u}(n)}}. \end{multline}

Next, to tackle the remaining nonconvexity of (\ref{P1.3_C54}), we have the following theorem:\\
\textbf{Theorem 1.}~\textit{The obtained solution $\boldsymbol{\omega}^{*}$ to the subproblem (\ref{opt:P1.3}) without the constraint (\ref{P1.3_C54}) can satisfy $\textrm{rank}~(\mathbf{W}_{b,k})=1$ \cite{zuo2022joint}, \cite{mu2020exploiting}.}\\
\begin{IEEEproof}
Problem (\ref{opt:P1.4}) is a convex problem. Therefore, the optimal solution is characterized by the KKT conditions. Specifically, the Lagrangian function of Problem (\ref{opt:P1.4}) in terms of the active beamforming $\bold{W}_{b,k}$ can be written as
\begin{multline}
|\hat{h}_{b,v}\omega_{b,k}|^{2}p_b \left( {I^{b,\textrm{intra}}_{k,u} + I^{b,\textrm{inter}}_{k,u} + \sigma^2} \right) \\ \geq |\hat{h}_{b,u}\omega_{b,k}|^{2}p_b \left( {I^{b,\textrm{intra}}_{k,v \rightarrow u} + I^{b,\textrm{inter}}_{k,v \rightarrow u} + \sigma^2} \right),
\label{B_1}\end{multline}
where $\bold{T}$ denotes the terms which are independent of $\bold{W}_{b,k}$. $\varsigma_{ku}, \tau_{ku}, \alpha$ and $\bold{Y}_{b,k}$ are lagrange multipliers associated with the corresponding constraints (\ref{P1.3_C2}), (\ref{P1.3_C3}), (\ref{P1.3_C4}), and (\ref{P1.3_C6}). The KKT conditions for the optimal $\bold{W}^{*}_{b,k}$ can be:
\begin{equation}
\varsigma^{*}_{ku}, \tau^{*}_{ku}, \alpha^{*} \geq 0, \bold{Y}^{*}_{b,k} \succeq \bold{0}, \bold{Y}^{*}_{b,k}\bold{W}^{*}_{b,k} =  \bold{0}, \nabla_{\bold{W}^{*}_{b,k}} \mathcal{L} =0,
\label{B_2}\end{equation}
where $\varsigma^{*}_{ku}, \tau^{*}_{ku}, \alpha^{*} $ and $\bold{Y}^{*}_{b,k}$ denote the optimal Lagrange multipliers. Then, the condition $\nabla_{\bold{W}^{*}_{b,k}} \mathcal{L} = 0$ can be expressed as
\begin{multline}
\alpha^{*}\mathcal{I}=\bold{Y}^{*}_{b,k}+\sum_{u\in C^{b}_{k}} \left(\varsigma^{*}_{ku}p_{b} - \tau^{*}_{ku} \sum^{U}_{\delta_{b,k}(u') > \delta_{b,k}(u)} p_{b}\right){\bold{H}^{H}_{bku}} \\
- \sum^{K_{b}}_{k' \ne k} \sum_{u \in C^{b}_{k}} \tau^{*}_{k'u}{\bold{H}^{H}_{bk'u}}.
\label{B_3}\end{multline}

It is observed that the matrix $\bold{Y}^{*}_{b,k}$ is a positive semidefinite matrix \cite{xu2020resource}. Therefore, we have $\textrm{rank}(\bold{Y}^{*}_{b,k}) = N_{T}-1$. Furthermore, if we assume that $\bold{Y}^{*}_{b,k}\bold{W}^{*}_{b,k} =  \bold{0}$, we can deduce $\textrm{rank}(\bold{W}^{*}_{b,k}) + \textrm{rank}(\bold{Y}^{*}_{b,k}) \leq N_{T}$. Finally, we can inferred that $\textrm{rank}(\bold{W}^{*}_{b,k})=1 $, because that $\bold{W}^{*}_{b,k}=\bold{0}$ is contradict with the minimum rate requirement constraints of problem (\ref{opt:P1.4}).
\end{IEEEproof} 

Based on Theorem 1, we can always obtain a rank one solution by solving the subproblem (\ref{opt:P1.3}) by ignoring the constraint (\ref{P1.3_C54}). As a result, the relaxed problem is a convex semidefinite program (SDP), which can be efficiently solved via standard convex problem solvers such as CVX \cite{diamond2016cvxpy}.

\subsubsection{Passive Beamforming based on Convex Approximation}
\label{sol_phase}
Similar to Section \ref{sol_power}, given the user association, pairing, and the active beamforming vector in problem Section.(\ref{opt:P1.2}), the passive beamforming optimization subproblem for any AP $b$ with the slack variables $\left\{ \nu^{b}_{k,u} |k \in \mathcal{K}_{b}, u \in \mathcal{U}_{b} \right\}$ and $\left\{\xi^{b}_{k,u}|k \in \mathcal{K}_{b}, u \in \mathcal{U}_{b} \right\}$ can be given by

\begin{maxi!}[2]<b> 
		{\substack{\boldsymbol{\Phi}, \left\{\nu^{b}_{k,u}, \xi^{b}_{k,u}, R^{b}_{k,u} \right\}}}   
		{\sum_{k=1}^{K_{b}} \sum_{u=1}^{U}  R^{b}_{k,u}}{\label{opt:P1.5}}{\textbf{P1.5:}}
        \addConstraint{[\mathbf{B}^{p}_{l}]_{m,m}=\beta^{p}_{m} \label{P1.4_C1}}
        \addConstraint{\mathbf{B}^{p}_{l} \succcurlyeq 0 \label{P1.4_C2}}
        \addConstraint{\textrm{rank}(\mathbf{B}^{p}_{l})=1 \label{P1.4_C3}}
		\addConstraint{\textcolor{black}{(\ref{P1_C7}),(\ref{P1.2_C1}),(\ref{P1.3_C1})\sim(\ref{P1.3_C3}),}}
\end{maxi!}
where $~\forall l \in \mathcal{L},~\forall m \in \mathcal{M},~\mathbf{B}^{p}_{l}=\Phi^{p}_{l}\Phi^{p}_{l},~p\in \left\{F,B\right\}$. Note that problem (\ref{opt:P1.5}) is nonconvex due to the rank-one constraint (\ref{P1.4_C3}). To tackle this issue, we apply semidefinite relaxation (SDR) by relaxing this constraint (\ref{P1.4_C3}) \cite{luo2010semidefinite}. Then, we can efficiently solve the problem (\ref{opt:P1.5}) without rank one constraint (\ref{P1.4_C3}) by standard convex optimization solvers such as CVX \cite{diamond2016cvxpy}. Therefore, by solving subproblems (\ref{opt:P1.4}) and (\ref{opt:P1.5}), the active beamforming vector $\boldsymbol{\omega}$ and the passive beamforming vector $\boldsymbol{\Phi}$ are alternately optimized. The solutions obtained after each iteration are used as the input local points for the following iteration. Algorithm \ref{alg:beamforming} summarizes the proposed successive convex approximation (SCA)-based algorithm to solve the problem (\ref{opt:P1.2}). It is noted that Algorithm \ref{alg:beamforming} is guaranteed to converge to a locally optimal solution of the subproblem (\ref{opt:P1.2}) \cite{dinh2010local}. However, we notice that the solution found using Algorithm \ref{alg:beamforming} is not a global optimal. This is because in Algorithm \ref{alg:beamforming}, we replaced non-convex constraints with approximate formulas to solve the problem (\ref{opt:P1.2}) through convex approximation. Therefore, we explain in the next section how to use this suboptimal obtained through Algorithm \ref{alg:beamforming} as mimic values for reinforcement learning.

\begin{algorithm}[t]
	\caption{\strut Successive Convex Approximation (SCA) for the beamforming vectors optimization} 
	\label{alg:beamforming}
	\begin{algorithmic}[1]
        \IF{$b \in \mathcal{B}$}
	    \STATE{\textbf{Input:} Initial slack variables $\zeta^{b}_{k,u}(0)$, $\eta^{b}_{k,u}(0), \xi^{b}_{k,u}(0)$, $\nu^{b}_{k,u}(0)$, and Initialize the initial vectors $\omega^{(0)}_b, \Phi^{(0)}_b$, and set the iteration index $n=0$.}
        \STATE{$\Xi^{(0)} = \sum_{k=1}^{K_{b}} \sum_{u=1}^{U} R^{b}_{k,u}(\omega^{(0)}, \Phi^{(0)})$}
        \REPEAT
        \STATE{update $\zeta^{b}_{k,u}(n+1)$, $\eta^{b}_{k,u}(n+1)$ and $\omega^{(n+1)}_b$ by solving problem (\ref{opt:P1.3})}
        \STATE{update $\nu^{b}_{k,u}(n+1)$, $\xi^{b}_{k,u}(n+1)$ and $\Phi^{(n+1)}_b$ by solving problem (\ref{opt:P1.4})}
        \STATE{$\Xi^{(n+1)} = \sum_{k=1}^{K_{b}} \sum_{u=1}^{U} R^{b}_{k,u}(\omega^{(n+1)}_b, \Phi^{(n+1)}_b)$}
        \STATE{$n=n+1$}
        \UNTIL{ $\left|\Xi^{(n+1)} - \Xi^{(n)} \right| \leq \epsilon$}
        \ENDIF
		\STATE{\textbf{Output:} The active beamforming vector $\boldsymbol{\omega}^*$ and the passive beamforming vector $\boldsymbol{\Phi}^*$.}
	\end{algorithmic}
\end{algorithm}

\subsubsection{Convex Approximation Imitated-based DRL}
\label{sol_camappo}

As presented in Fig. \ref{fig_camappo}, we propose a proximal policy optimization (PPO)-based MARL to solve problem $\mathbf{P1.2}$. PPO is a widely used reinforcement learning technique that is known for its simplicity in implementation and applicability across diverse situations \cite{ppo}. Furthermore, it has demonstrated consistent and reliable performance. The utilization of the Proximal Policy Optimization (PPO) technique serves to streamline the intricate computational process associated with TRPO. TRPO algorithm aims to optimize a surrogate objective function in the following manner \cite{trpo}.

\begin{equation}
L^{\textrm{TRPO}}_{n}(\phi)=\hat{\mathbb{E}}_{n} \left[ {\frac{\pi_{\phi}(\mathcal{A}_{n}|\mathcal{S}_{n})}{\pi_{\phi_{\textrm{old}}}(\mathcal{A}_{n}|\mathcal{S}_{n})}} \hat{A}_{n}  \right] = \hat{\mathbb{E}}_{n} \left[ r_{n}(\phi)\hat{A}_{n} \right],\label{L_TRPO}
\end{equation}
where $r_{n}(\phi)$ denotes the probability ratio, $\mathcal{A}_{n}$ and $\mathcal{S}_{n}$ are an action and reward in time step $n$. The surrogate objective function $L^{\textrm{TRPO}}$ of TRPO has a complicated formula expansion and must calculate the second derivative. Therefore, maximization of the surrogate objective function $L^{\textrm{TRPO}}$ of TRPO would result in an unnecessarily large policy update in the absence of a constraint. Therefore, in PPO, the limitations of TRPO were addressed by incorporating an approximation of the first derivative using the clipping technique. The following is the objective function to which the clipping is applied:
\begin{equation}
L^{\textrm{CLIP}}_{n}(\phi)=\hat{\mathbb{E}}_{n} \left[ \textrm{min}(r_{n}(\phi)\hat{A}_{n}, \textrm{clip}(r_{n}(\phi), 1-\epsilon, 1+\epsilon)\hat{A}_{n}) \right],\label{L_CLIP}
\end{equation}
where $\epsilon$ is a hyperparameter and $\hat{A}_{n}$ is a truncated version of generalized advantage estimation which can be defined as follows:
\begin{equation}
    \hat{A}_{n} = \delta_{n} + (\gamma\lambda)\delta_{n+1} + \dots + (\gamma\lambda)^{N-n+1}\delta_{N-1},
\label{eq_adv}
\end{equation}
where $\delta_{n}=r_{n}+\gamma V(s_{n+1}) - V(s_{n})$. The function in \eqref{L_CLIP} takes a lower value when comparing the objectives used in the TRPO with the objectives to which clipping is applied. With this clipping method, we only consider the change in probability ratio if it improves the objective. If it makes the objective worse, we leave it out.

Subsequently, PPO incorporates an actor-critic network architecture, wherein the policy and value functions share the parameters inside the network design. In the context of utilizing a neural network architecture to share parameters between the policy function and the value function, it is imperative to implement a loss function that effectively integrates the policy surrogate and an error term derived from the value function. The objective is achieved by integrating the policy surrogate with a value function error component as follows:
\begin{equation}
L^{\textrm{PPO}}_{n}(\phi)=\hat{\mathbb{E}}_{n} \left[ L^{\textrm{CLIP}}_{n}(\phi) - c_{1} L^{\textrm{VF}}_{n}(\phi) + c_{2}E[\pi_\phi](s_n) \right],\label{L_PPO}
\end{equation}
where $c_1$ and $c_2$ are coefficients, $E$ denotes an entropy function, and $L^{\textrm{VF}}_{n}$ is a squared-error loss. In this objective $L^{\textrm{PPO}}$ can be further increased by adding an entropy bonus $E$ to ensure sufficient exploration.

In addition, we propose to imitate the locally optimal solutions derived through Algorithm \ref{alg:beamforming} as loss during the learning process to ensure convergence and maximize speed in the early learning process. Therefore, the difference between the action derived from the PPO model and the locally optimal solution derived through Algorithm \ref{alg:beamforming} can be defined as follows:

\begin{equation}
L^{\textrm{CA}}_{n}(\phi)= \frac{1}{n_a} \sum^{n_a}_{i=1} (\tilde{\mathcal{A}}_{n}(i) - \mathcal{A}_{n}(i))^{2},\label{L_CA}
\end{equation}
where $n_a$ is the number of action space, $\tilde{\mathcal{A}}_{n}$ is the locally optimal solution derived through Algorithm \ref{alg:beamforming}, and $ \mathcal{A}_{n}$ is action determined by the policy $\phi$ being learned. Finally, we define the loss of the following learning model as follows:

\begin{equation}
L^{\textrm{TOTAL}}_{n}(\phi)= L^{\textrm{PPO}}_{n}(\phi) + \psi \cdot L^{\textrm{CA}}_{n}(\phi),\label{L_TOTAL}
\end{equation}
where $\psi$ is the imitation rate. We propose that the imitation rate $\psi$ is gradually reduced as learning proceeds from the initial value, thus, learning can find the final optimal value. In the simulations, we show the learning result according to the change in the imitation rate and to the initial imitation rate.

\begin{algorithm}[!t]
	\caption{\strut Convex approximation imitated-based multi-agent PPO (CAMAPPO)} 
	\label{alg:CAMAPPO}
	\begin{algorithmic}[1]
	    \STATE{\textbf{Initialize:} the initial network $\phi^{\omega}_{0}$ and $\phi^{\Phi}_{0}$ for agent of active and passive beamforming coefficients, respectively.}
		\FOR{episode$=1,2,...,E$}
		\STATE{Initialize randomly each UE's position and excute 2-Stage Matching for UE Association and Pairing from Section \ref{sol_3dmatching}, i.e., $\boldsymbol{\alpha}^{*}, \boldsymbol{\gamma}^{*}$.}
		\FOR{time slot$=1,2,...,N$}
		\STATE{Update observation $\mathcal{S}(n)$.}
		\STATE{Run policy $\mathcal{A}^{\omega}\sim\pi^{\omega}_{\phi_{\textrm{old}}}, \mathcal{A}^{\Phi}\sim\pi^{\Phi}_{\phi_{\textrm{old}}}$.}
		\STATE{Compute the common reward $\mathcal{R}(n)$.}
		\STATE{Save $(\mathcal{S}(n),\mathcal{A}(n),\mathcal{R}(n),\mathcal{S}(n+1))$ in memory of each agent.}
		\ENDFOR
		\STATE{Compute advantage estimates $\left\langle \hat{A}^{\omega}_{1},...,\hat{A}^{\omega}_{N} \right\rangle$, $\left\langle \hat{A}^{\Phi}_{1},...,\hat{A}^{\Phi}_{N} \right\rangle$  based on (\ref{eq_adv}).}
		\STATE{Optimize surrogate $L^{\textrm{TOTAL}}$ wrt $\phi^{\omega}$, $\phi^{\Phi}$ with minibatch from memory based on (\ref{L_TOTAL}).}
		\STATE{$\phi^{\omega}_{\textrm{old}} \leftarrow \phi^{\omega}$, $\phi^{\Phi}_{\textrm{old}} \leftarrow \phi^{\Phi}$}
		\ENDFOR
		\STATE{\textbf{Output:} The proposed networks $\pi^{\omega}_{\phi_{\textrm{opt}}}$, $\pi^{\Phi}_{\phi_{\textrm{opt}}}$.}
	\end{algorithmic}
\end{algorithm}

The proposed convex approximation imitated-based multi-agent PPO (CAMAPPO) in this paper is for each optimization variable ($\boldsymbol{\omega}, \boldsymbol{\Phi}$) PPO agents. Each agent learns simultaneously in the same environment. Moreover, each agent can determine the optimal action for a common reward. Therefore, we introduce the Markov Decision Process (MDP) of the agents used for learning.
\begin{equation}
\mathcal{S}(n) =  \{\boldsymbol{\alpha}^{*}, \boldsymbol{\gamma}^{*}, \boldsymbol{\omega}(n-1), \boldsymbol{\Phi}(n-1), \{ \hat{h}_{u} \}_{u \in \mathcal{U}} \} \label{eq_state},
\end{equation}
\begin{equation}
\mathcal{A}^{\omega}(n)=\{\omega_{b}(n)\}_{b\in \mathcal{B}},
\end{equation}
\begin{equation}
\mathcal{A}^{\Phi}(n)=\{\Phi^{F}_{l}, \Phi^{B}_{l}\}_{l\in \mathcal{L}},
\end{equation}
\begin{equation}
\mathcal{R}(n)= \textrm{min}_{u \in \mathcal{U}}(R_{k,u}),
\end{equation}
where $\boldsymbol{\omega}(n-1)$ and $\boldsymbol{\Phi}(n-1)$ denote the active and passive beamforming vector at the last step, $\mathcal{A}^{\omega}(n)$ and $\mathcal{A}^{\Phi}(n)$ denote the active and passive beamforming agent. The reward is determined by the lowest throughput of all users. This is to satisfy the minimum throughput of all users for the constraint (\ref{P1_C1}) in problem (\ref{opt:P1.1}). Based on the MDP for the proposed optimization problem, the solution of the decision variable can be obtained by executing the proposed CAMAPPO algorithm, such as Algorithm \ref{alg:CAMAPPO}. \textcolor{black}{For the proposed RL approach, its computational complexity depends mainly on the size of the neural network. Therefore, its computational complexity is about $\mathcal{O}(\sum^{L}_{l=1}n_{l-1}\cdot n_{l})$, where $L$ is the number of network layers and $n_l$ is the number of neurons in the $l$-th layer.}

\begin{algorithm}[t]
	\caption{\strut Joint block coordinate descent(BCD)} 
	\label{alg:bcd}
	\begin{algorithmic}[1]
        \STATE{\textbf{Initialize:} $n=-1$, the proposed network $\pi^{\omega}_{\phi_{\textrm{opt}}}$, $\pi^{\Phi}_{\phi_{\textrm{opt}}}$ and $\boldsymbol{\alpha}^{(0)}, \boldsymbol{\gamma}^{(0)}, \boldsymbol{\delta}^{(0)}, \boldsymbol{\omega}^{(0)}, \boldsymbol{\Phi}^{(0)}$.}
        \STATE{$\Xi^{(0)} = \sum_{k=1}^{K_{b}} \sum_{u=1}^{U} R^{b}_{k,u}(\boldsymbol{\alpha}^{(0)}, \boldsymbol{\gamma}^{(0)}, \boldsymbol{\delta}^{(0)}, \boldsymbol{\omega}^{(0)}, \boldsymbol{\Phi}^{(0)})$}
        \REPEAT
        \STATE{$n\leftarrow n+1$}
        \STATE{update $\boldsymbol{\alpha}^{(n+1)}$ for a given $\boldsymbol{\gamma}^{(n)}, \boldsymbol{\delta}^{(n)}, \boldsymbol{\omega}^{(n)}, \boldsymbol{\Phi}^{(n)}$ according to Algorithm \ref{alg:association};}
        \STATE{update $\boldsymbol{\gamma}^{(n+1)}$ for a given $\boldsymbol{\alpha}^{(n)}, \boldsymbol{\delta}^{(n)}, \boldsymbol{\omega}^{(n)}, \boldsymbol{\Phi}^{(n)}$ according to Algorithm \ref{alg:pairing};}
        \STATE{update $\boldsymbol{\delta}^{(n+1)}$ for a given $\boldsymbol{\alpha}^{(n)}, \boldsymbol{\gamma}^{(n)}, \boldsymbol{\omega}^{(n)}, \boldsymbol{\Phi}^{(n)}$ according to Section \ref{sol_decoding};}
        \STATE{update $\boldsymbol{\omega}^{(n+1)}, \boldsymbol{\Phi}^{(n+1)}$ for a given $\boldsymbol{\alpha}^{(n)}, \boldsymbol{\gamma}^{(n)}, \boldsymbol{\delta}^{(n)}$, $ \boldsymbol{\omega}^{(n)}, \boldsymbol{\Phi}^{(n)}$ by running network $\pi^{\omega}_{\phi_{\textrm{opt}}}, \pi^{\Phi}_{\phi_{\textrm{opt}}}$, respectively;}        
        \STATE{$\Xi^{(n+1)} = \sum_{k=1}^{K_{b}} \sum_{u=1}^{U} R^{b}_{k,u}$;}
        \UNTIL{ $\left|\Xi^{(n+1)} - \Xi^{(n)} \right| \leq \epsilon$}
		\STATE{\textbf{Output:} The proposed solution sets $\boldsymbol{\alpha}^{(n+1)}, \boldsymbol{\gamma}^{(n+1)}, \boldsymbol{\delta}^{(n+1)}$, $ \boldsymbol{\omega}^{(n+1)}, \boldsymbol{\Phi}^{(n+1)}$ .}
	\end{algorithmic}
\end{algorithm}

\textcolor{black}{
\subsection{Convergence and Complexity Analysis}
After obtaining a model that completed the training through Algorithm \ref{alg:bcd}, we apply block coordinate decision (BCD) technology to find the proposed solution \cite{tseng2001convergence}. To address the interdependence between decision variables within objective functions and constraints, it is supported by iterative solutions referenced in \cite{lyu2020block} and \cite{hong2017iteration}, strategies that manage computational complexity inherent in MINLP optimization by decomposing the main problem into three subproblems. Furthermore, our devised BCD algorithm can achieve $\epsilon$-optimal solutions with sub-linear convergence rates denoted by $\mathcal{O}(1/n)$, as explicitly presented in Algorithm \ref{alg:bcd}. The comprehensive approach results in an aggregate computational complexity of the overall BCD Algorithm \ref{alg:bcd} is summarized as $\mathcal{O}(\mathcal{B}\mathcal{N}^{\textrm{bcd}}(\mathcal{U} + \mathcal{N}^{\textrm{pair}}(\mathcal{K}^{2}\mathcal{U}+\mathcal{K}^{2}+\mathcal{K}\mathcal{M}) + \sum^{L}_{l=1}n_{l-1}\cdot n_{l})$, where $\mathcal{N}^{\textrm{bcd}}$ and $\mathcal{N}^{\textrm{pair}}$ represent iteration of BCD algorithm and UE pairing, respectively.
}

\begin{figure*}[!t]
        \centering
        \begin{subfigure}[t]{0.29\textwidth}
        \centering
               \captionsetup{justification=raggedright,singlelinecheck=false}
                \includegraphics[width=\linewidth]{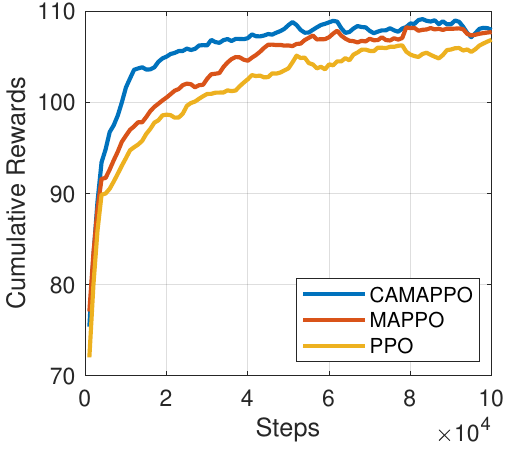}
                \caption{Learning convergence results.}
                \label{simulation_conv1}
        \end{subfigure}%
        \begin{subfigure}[t]{0.29\textwidth}
        \centering
               \captionsetup{justification=raggedright,singlelinecheck=false}
               \includegraphics[width=\linewidth]{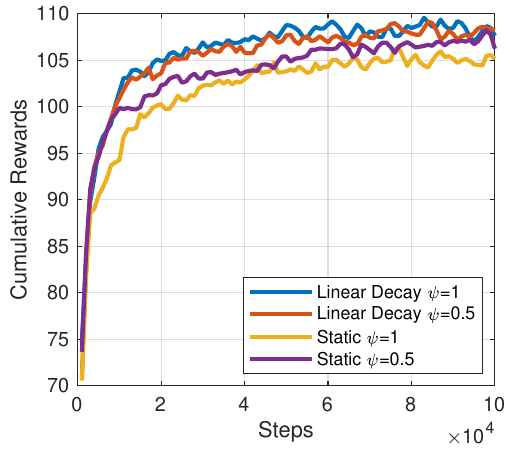}
                \caption{Convergence w.r.t imitation rate $\psi$.}
                \label{simulation_conv2}
        \end{subfigure}%
        \begin{subfigure}[t]{0.29\textwidth}
        \centering
         \captionsetup{justification=raggedright,singlelinecheck=false}
         \includegraphics[width=\linewidth]{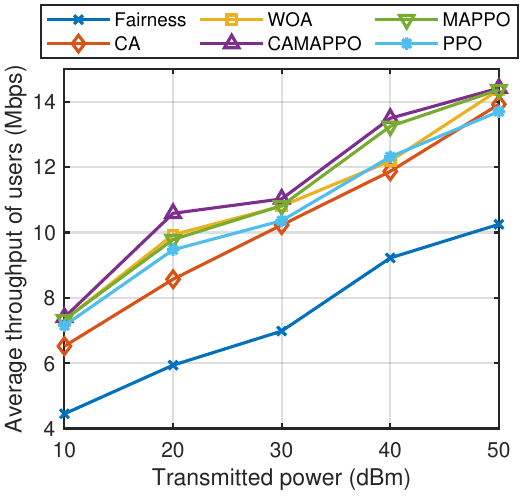}
                \caption{Throughput of UEs vs power.}
                \label{simulation_avgDR}
        \end{subfigure}%
        \caption{Analysis of learning convergence and performance with different algorithms.}\label{mixed2}
\end{figure*}
\begin{figure}[!t]
    \centering
    \includegraphics[width=0.55\columnwidth]{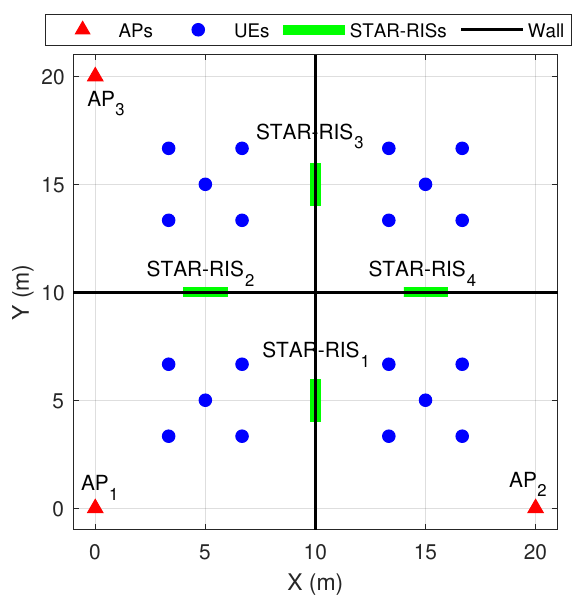}
    \caption{\textcolor{black}{Deployments of AP, STAR-RISs, UEs.}}
    \label{simulation_deploy}
\end{figure}

\section{Simulation Settings and Results}
\label{simul}
\begin{figure*}[t]
    \centering
    \includegraphics[width=0.8\textwidth, height=1.8in]{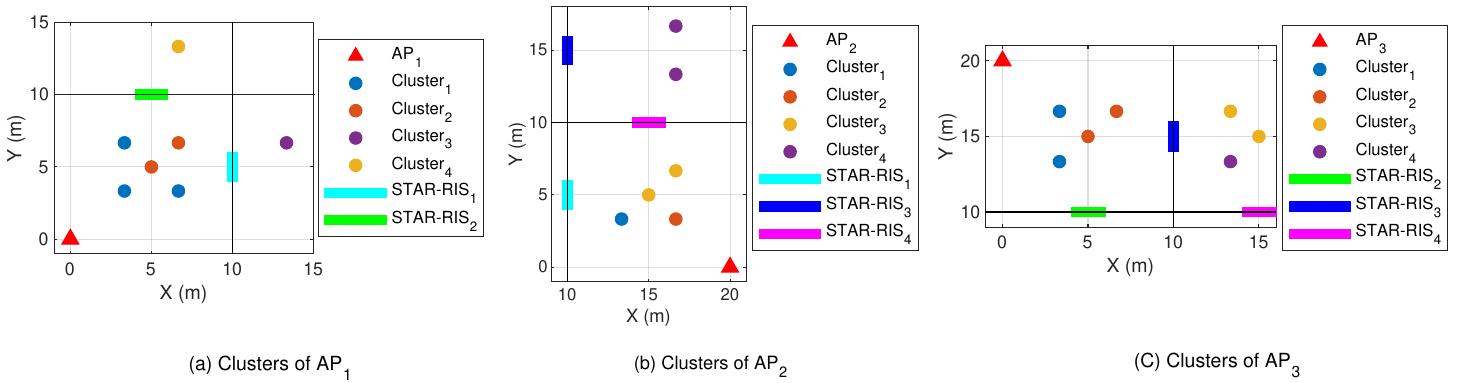}
    \caption{Results of clustering for each AP after 2-Stage Matching.}
    \label{simulation_cluster}
\end{figure*}

We consider a region of $20~$m$^2$ served by three APs catering to $20$ UEs. This region is further divided into four rooms, each enclosed by walls and equipped with four Star-RIS units. Each AP implements a NOMA network with four clusters. Initially, UEs are randomly distributed across the rooms during the learning stage. For performance verification as shown in Fig.~\ref{simulation_cluster}, UEs are assigned to specific clusters.
The APs are equipped with four antennas operating at $6$ GHz with a noise density of $-100$ dBm/Hz. A Rician fading factor of $4$ dB is adopted. By considering the path loss model for the indoor hotspot scenario, as presented in 3GPP TR 38.901 version 16.1.0 Release 16, the path losses, at the reference distance of the LoS link $PL^{\textrm{LoS}} = 32.4+17.3\log_{10}{(d)}+20\log_{10}{(f)}$ and NLoS link $PL^{\textrm{NLoS}} = 32.4+31.9\log_{10}{(d)}+20\log_{10}{(f)}$, where $d$ is the distance between the transmitter and the receiver, and $f$ is the sub-carrier frequency. The channel between an AP and an UE can be modeled as a Rician channel, which includes one LoS path and several NLoS paths. Each Star-RIS consists of $26$ elements spaced $0.2$ units wide and $0.1$ units long. For learning purposes, an episode comprises $10$ time slots. The learning network model utilizes two hidden layers, each with $128$ units. A correlation-based K-means clustering algorithm is employed for UE pairing after the initial random assignment at the beginning of each episode.

The proposed CAMAPPO algorithm, a variant of MAPPO with a convex approximation, is compared with four other algorithms and their details are provided subsequently.
\begin{itemize}
    \item \emph{CAMAPPO (proposed)}: Our proposed method treats each beamforming vector variable as a separate learning agent within a MAPPO framework. To enhance learning efficiency, it incorporates a convex approximation (CA) approach.
    \item \emph{MAPPO}: The algorithm employs a decentralized approach, where each variable is optimized by a separate PPO agent, and convexity is not enforced during the learning process.
    \item \emph{PPO}: The general PPO approach performs a joint optimization of all beamforming vector variables through a single neural network.
    \item \emph{CA}: For each variable, the approach leverages a convex approximation technique to obtain a suboptimal solution, which does not consider the requirement for a dedicated learning component. 
    \item \emph{WOA}: The benchmark method utilized the whale optimization algorithm, a meta-heuristic technique, to find the sub-optimal value for each variable.
\end{itemize}

\begin{figure}[!t]
    \centering
    \includegraphics[width=0.55\columnwidth]{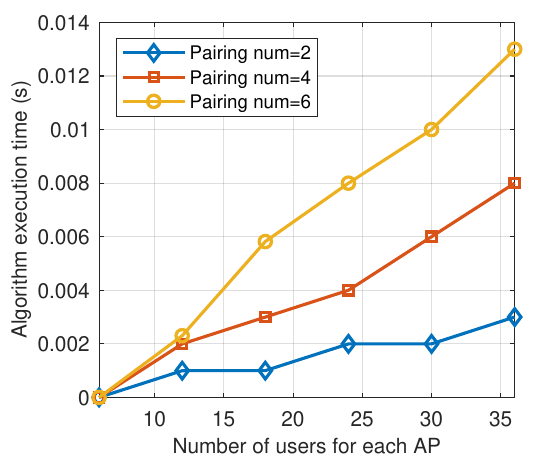}
    \caption{\textcolor{black}{Correlation-based K-means Clustering time depending on the number of users of the AP.}}
    \label{simulation_ap_excutiontime}
\end{figure}

\begin{figure*}[t]
    \centering
    \includegraphics[width=0.8\textwidth, height=2in]{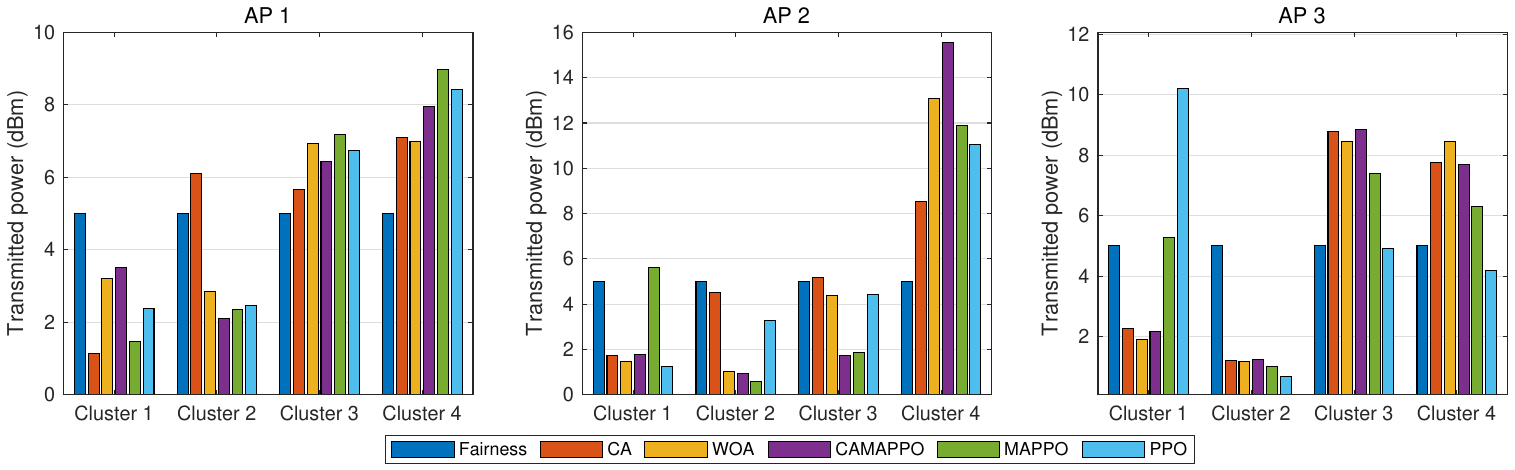}
    \caption{Status of active beamforming allocation for clusters by algorithms.}
    \label{simulation_active}
\end{figure*}

Fig.~\ref{simulation_conv1} depicts the convergence process of the compared reinforcement learning algorithms. Our proposed CAMAPPO algorithm achieved convergence within approximately $30,000$ steps, demonstrating significant improvement in learning speed. It converged roughly twice as fast as MAPPO and $2.3$ times faster than the baseline PPO algorithm. Interestingly, the final converged performance levels were similar for CAMAPPO and MAPPO, with both achieving a higher cumulative reward compared to PPO (approximately 1$\%$ lower). These results suggest that the imitation learning component within CAMAPPO plays a crucial role in accelerating the convergence process. Furthermore, the superior performance of MAPPO compared to PPO highlights its effectiveness in handling complex variable spaces by facilitating faster and more accurate learning.

Fig.~\ref{simulation_conv2} explores the convergence behavior of the algorithm regarding the imitation rate ($\psi$). As observed, excluding the static case ($\psi=0.5$), all scenarios achieved similar levels of cumulative reward at convergence. This suggests a potential influence of the imitation rate on learning performance. Although a higher imitation rate ($\psi$ closer to $1$) can be beneficial in the initial stages by guiding the learning agent, it likely needs to be gradually reduced ($\psi$ closer to $0$) during the later stages to steer the agent toward independently finding the optimal solution for the specific problem. Identifying the optimal dynamic adjustment strategy for the imitation rate is an important area for future research efforts.

\begin{figure*}[t]
    \centering
    \includegraphics[width=0.8\textwidth, height=2in]{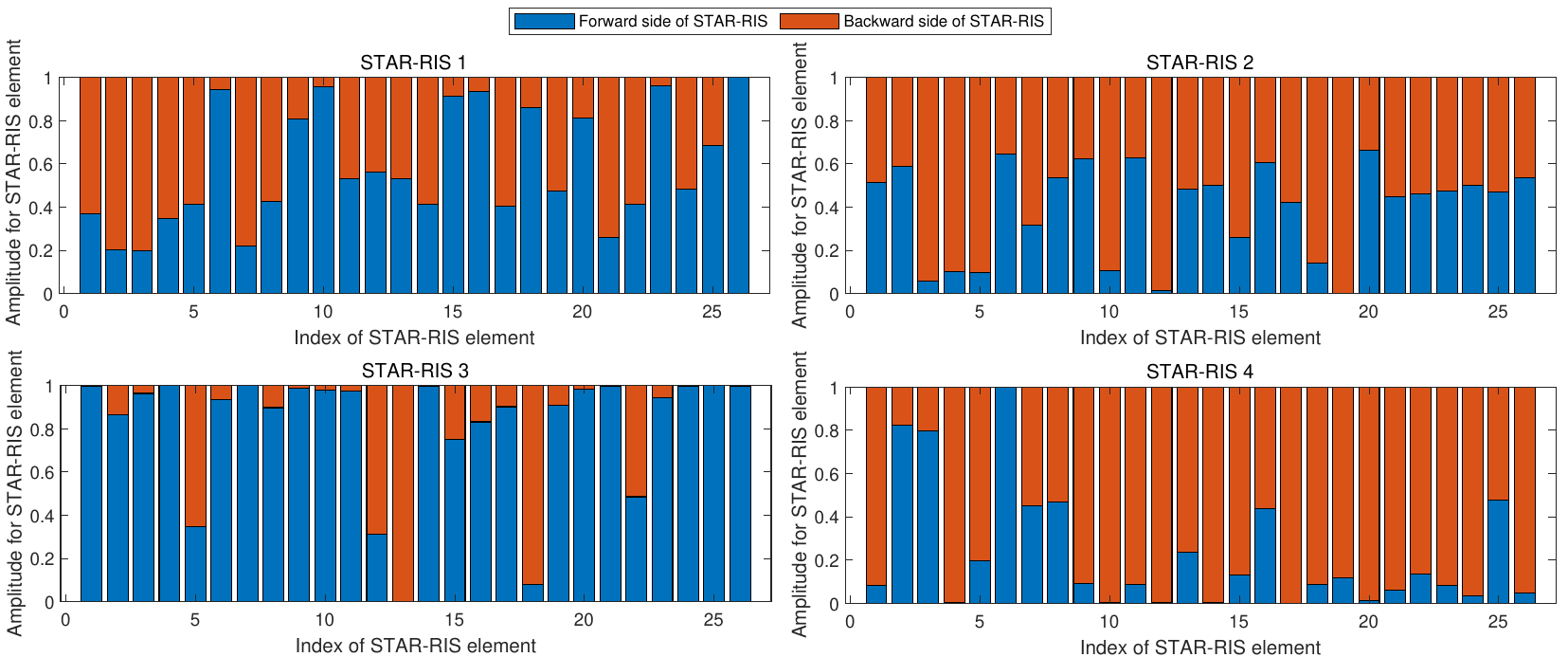}
    \caption{Amplitude of each element of STAR-RIS derived through the proposed algorithm (CAMAPPO).}
    \label{simulation_amplitude}
\end{figure*}

\textcolor{black}{Fig.~\ref{simulation_avgDR} represents the average UE throughput as a function of the transmission power for each AP. The proposed CAMAPPO algorithm consistently achieves higher average throughput compared to all other benchmark algorithms at all transmission power levels. Both CAMAPPO and MAPPO used the same neural network architecture, number of iterations, and other learning parameters, but MAPPO showed an average lower performance of 2.5$\%$. This can be due to the inherent limitations of MAPPO in complex learning scenarios, and additional learning iteration is potentially required to achieve optimal performance. As shown in Fig.~\ref{simulation_conv1}, PPO is a network model of the same size, and learning is inadequate in the process of optimizing all beamforming vectors. Therefore, the performance of PPO also showed about 7$\%$ lower performance on average when compared to CAMAPPO. It is noteworthy that the performance improvement of CAMAPPO is much more pronounced when compared to existing resource optimization algorithms such as CA and WOA. Compared to CAMAPPO, WOA showed an average performance of around 4$\%$. In particular, CAMAPPO outperforms CA by 10$\%$ and up to 15$\%$, on average. This gap arises from the inherent limitations of CA in converting the original non-convex optimization problem into a convex problem. This conversion process often leads to suboptimal solutions over the more flexible approaches used by CAMAPPO. In addition, in terms of solution inference speed, CA and WOA took 20s and 40s, respectively, but inference through network models took 0.002s under the same conditions. The proposed method has the additional advantage of deriving faster solutions even in real-world environments.}

Fig.~\ref{simulation_deploy} depicts the network topology employed in our simulation environment. This specific topology serves as the benchmark for evaluating the performance of our proposed algorithm against the chosen baseline algorithms. The simulated network consists of a $20$ square meter region covered by three APs responsible for serving $20$ UEs. To introduce environmental complexity, this region is further divided into four separate rooms, each enclosed by walls. Additionally, each room is equipped with four Star-RIS units. To account for dynamic user distribution, the initial placement of UEs within the rooms is randomized during the learning stage of the algorithms. This initial random placement reflects real-world scenarios where user locations can vary.

The clustering algorithm should be determined quickly to determine the communication optimization of the users indoors. Fig.~\ref{simulation_ap_excutiontime} shows the execution time of the algorithm for each available pairing number according to the number of users in clustering. A typical NOMA network considers two or four pairs, but in this experiment, up to six pairs were considered to assume a harsh communication environment. As a result, for users within 40, clustering could be determined with a time lower than 0.014 seconds in all pairings. It took a while. Therefore, it can be determined that the proposed algorithm can be applied to the communication optimization of an actual indoor environment.

Fig.~\ref{simulation_cluster} shows UE clustering results obtained after each AP completes the two-stage matching process. This matching process employs correlation-based K-means clustering to group UEs with similar channel correlations. As evident in the figure, the proposed algorithm effectively assigns UEs to different clusters based on their physical locations (different rooms) and channel characteristics. This spatial separation is achieved by leveraging the correlation information within the clustering algorithm. Notably, the proposed method successfully pairs UEs situated in different rooms with the corresponding APs, promoting efficient resource allocation and potentially improved network performance, which validates our proposal.

Fig.~\ref{simulation_active} illustrates the transmission power allocated by each algorithm to the APs depicted in the communication environment as shown in Fig.~\ref{simulation_cluster}. For AP $1$, clusters $3$ and $4$ reside in a different room, limiting the potential for increased overall throughput simply by raising the transmission power. To address this, CAMAPPO strategically allocates higher transmission power to clusters $3$ and $4$, aiming to compensate for the distance-related attenuation. Additionally, cluster $1$, which has three UEs assigned, receives more power compared to cluster $2$ with only one UE. This prioritization reflects the need to cater to a larger number of users in cluster $1$. In contrast, algorithms like CA and Fairness, which exhibited lower performance in Fig.~\ref{simulation_avgDR}, likely allocated transmission power unevenly, as indicated by the irregular power distribution in the figure. This inefficient allocation likely contributed to their lower overall throughput. Similar trends are observed for AP $2$ and AP $3$. By allocating higher transmission power to UEs located in the same room as the corresponding AP, CAMAPPO prioritizes those experiencing less path loss, ultimately maximizing the final average throughput. This analysis highlights the effectiveness of CAMAPPO's dynamic power allocation strategy in enhancing network performance compared to benchmark algorithms.

Fig.~\ref{simulation_amplitude} presents the element-wise amplitudes for all STAR-RIS units as determined by the proposed CAMAPPO algorithm. To gain deeper insights, we analyze the sum of amplitudes across each element. For STAR-RIS 1, the amplitude distribution exhibits a clear bias toward the forward direction with a ratio of $15.141:10.858$, compared to the backward direction. Similarly, STAR-RIS 2 shows a forward bias with a ratio of $10.201:15.798$. Examining these results in conjunction with the network topology (Fig.~\ref{simulation_deploy}), we can infer that this prioritization of the forward direction aims to enhance throughput for Clusters 3 and 4 associated with AP 1. These clusters are located in different rooms compared to AP 1, making forward direction amplification crucial for improved signal strength. The amplitude distribution for STAR-RIS 3 and 4 reveals a contrasting pattern. Here, we observe ratios of $21.138:4.861$ and $5.861:20.138$, respectively. This significant bias towards reflection can be attributed to the absence of an AP in the first quadrant for these STAR-RIS units (refer to Fig.~\ref{simulation_deploy}). Consequently, the proposed algorithm prioritizes reflecting the signal rather than transmitting directly, effectively utilizing the STAR-RIS units to manage signal propagation within the environment. The analysis of STAR-RIS element amplitudes provides valuable insights into how CAMAPPO dynamically adjusts these elements to optimize network performance within the specific experimental environment.

\begin{table}[t]
    \centering
	\caption{Inference seconds time to optimize beamforming according to the number of users in the AP}
	\setlength{\tabcolsep}{2.6pt}
	{\footnotesize
		\renewcommand{\arraystretch}{1.2}
		\begin{tabular}{|p{55pt}| p{25pt}| p{25pt}| p{25pt}| p{25pt}| p{25pt}|}
			\hline
			\textbf{Algo./Num.}  & \textbf{18} & \textbf{22}& \textbf{26}& \textbf{30}& \textbf{34}	\\
			\hline
			\hline
			CA & 15.059 & 18.141 & 21.095 & 24.410 & 26.998\\
			\hline
			WOA & 39.073 & 40.531 & 41.646 & 42.697 & 45.031\\
			\hline
			\textbf{CAMAPPO} & 0.002 & 0.002 & 0.002 & 0.002 & 0.002\\
			\hline
			MAPPO & 0.002 & 0.002 & 0.002 & 0.002 & 0.002\\
			\hline
			PPO & 0.001 & 0.001 & 0.001 & 0.001 & 0.001\\
			\hline
		\end{tabular}
	}
	\label{tab1_time}
\end{table}

In order to apply the algorithm proposed in a real-world environment, it is essential to verify the inference time. Table \ref{tab1_time} shows the optimal inference time for each algorithm according to the number of users. It can be seen that the inference time of CA and WOA, which were used for general optimization, linearly increases as the number of users increases. However, reinforcement learning-based methods were robust to the number of users and were able to infer quickly. Therefore, the reinforcement learning-based method proposed in an actual communication environment where it is important to quickly derive the optimal value can be applied. However, the actual learning time was faster in general reinforcement learning methods because learning time requires additional time to derive suboptimal values. To solve this problem in future work, data on suboptimal values will be created before learning to save time to derive suboptimal values during learning.

\section{Conclusion}
\label{conc}
This work proposes a novel network architecture for indoor wireless communication that seamlessly integrates multiple APs with STAR-RIS units. This architecture aims to deliver exceptional service to a multitude of UEs. We depart from traditional approaches by formulating a comprehensive optimization problem that tackles UE assignment, AP beamforming, and dynamic STAR-RIS control within a unified framework. To address the inherent complexity arising from heterogeneous decision variables with diverse constraints, we introduce a decentralized DRL framework that empowers each decision variable to act as an independent agent, enabling them to learn and adapt within the dynamic environment. Furthermore, we present CAMAPPO, a novel solution approach that incorporates learned values and leverages a CA for significantly faster learning and improved convergence. Extensive simulation results demonstrate the superiority of CAMAPPO which outperforms baseline algorithms by a significant margin in terms of network utility, solidifying its exceptional performance within the specified network environment.

\ifCLASSOPTIONcaptionsoff
  \newpage
\fi
\bibliographystyle{IEEEtran}
\bibliography{ref}

\end{document}